\documentclass{aa}  
\usepackage{graphicx}
\usepackage{txfonts}
\usepackage{lipsum}
\usepackage{subcaption}
\usepackage{amssymb}

\begin{document} 
    \title{The ROAD to discovery: machine learning-driven anomaly detection in radio astronomy spectrograms}
\titlerunning{Machine learning anomaly detection in radio astronomy spectrograms}

\author{M. Mesarcik \inst{1}
      \and
      A. J. Boonstra \inst{3}
      \and 
      M. Iacobelli\inst{3}
      \and 
      E. Ranguelova \inst{2}
      \and
      C. T. A. M. de Laat \inst{1}
      \and 
      R. V. van Nieuwpoort \inst{1,2}
      }
\authorrunning{M. Mesarcik et al.}

\institute{Informatics Institute, University of Amsterdam, Science Park 900, 1098 XH Amsterdam, The Netherlands\\ \email{m.mesarcik@uva.nl}
    \and
       Netherlands eScience Center, Science Park 402, 1098 XH Amsterdam, The Netherlands
    \and
        ASTRON, the Netherlands Institute for Radio Astronomy, Oude Hoogeveensedijk 4, 7991 PD Dwingeloo, The Netherlands 
         }

\date{Received September 15, 1996; accepted March 16, 1997}

\newcommand{\red}[9]{{\color{red}{#1}}}
\newcommand{\ROAD}[0]{{ROAD}}

  \abstract
   {As radio telescopes increase in sensitivity and flexibility, so do their complexity and data-rates. For this reason automated system health management approaches are becoming increasingly critical to ensure nominal telescope operations.}
   {We propose a new machine learning anomaly detection framework for classifying both commonly occurring anomalies in radio telescopes as well as detecting unknown rare anomalies that the system has potentially not yet seen. To evaluate our method, we present a dataset consisting of 7050 autocorrelation-based spectrograms from the Low Frequency Array (LOFAR) telescope and assign 10 different labels relating to the system-wide anomalies from the perspective of telescope operators. This includes electronic failures, miscalibration, solar storms, network and compute hardware errors among many more.}
   {We demonstrate how a novel Self Supervised Learning (SSL) paradigm, that utilises both context prediction and reconstruction losses, is effective in learning normal behaviour of the LOFAR telescope. We present the Radio Observatory Anomaly Detector (ROAD),  a framework that combines both SSL-based anomaly detection and a supervised classification, thereby enabling both classification of both commonly occurring anomalies and detection of unseen anomalies.}
   {We demonstrate that our system is real-time in the context of the LOFAR data processing pipeline, requiring <1ms to process a single spectrogram. Furthermore, ROAD obtains an anomaly detection F-2 score of 0.92 while maintaining a false positive rate of 2\%, as well as a mean per-class classification F-2 score 0.89, outperforming other related works.}
    {}

    \maketitle 
    \section{Introduction}
Radio telescopes are getting bigger and are generating increasing amounts of data to improve their sensitivity and resolution~\cite{Norris2010, VanHaarlem2013, Foley2016, Nan2011}. The growing system size and resulting complexity increases the likelihood of unexpected events occurring thereby resulting in datasets that contain anomalies. These anomalies include failures in instrument electronics, miscalibrated observations, environmental events such as lightning, astronomical effects like solar storms as well as problems in data processing systems among many others. We consider Radio Frequency Interference (RFI) unavoidable therefore do not consider it an anomaly in this context. Currently, efforts to detect and mitigate these anomalies are performed by human operators, who manually inspect intermediate data products to determine the success or failure of a given observation. The accelerating data-rates coupled with the lack of automation, results in operator-based data quality inspection becoming increasingly infeasible~\cite{Mesarcik2020b}. 

In the context of low-frequency radio astronomy, scientific data processing has been successfully automated by running complex workflows that perform calibration and imaging of interferometric data~\cite{Gasperin2019,Weeren2016,Tasse2018,Wijnholds2010}, Radio Frequency Interference (RFI) mitigation~\cite{Offringa2010a} and de-dispersion~\cite{Barsdell2012,Bassa2022} of time-domain data among many more. Additionally, continuous effort is being made to create high-performance real-time algorithms, to improve the quality and reliability of the scientific data~\cite{Sclocco2019,Sclocco2016,vanNiewpoort2011,LaPlante2021,Broekema2018}. However, as of yet, there have been no attempts to fully automate the System Health Management (SHM) pipeline, and by virtue of the lack of work on this topic, no real-time implementations exist. This is in part due to the complexity of the challenge as well as the unavailability of SHM specific datasets. Furthermore, the successes of SHM-based anomaly detection systems have been extremely impactful in fields ranging from industrial manufacturing~\cite{Bergmann2019a} to space craft system health~\cite{Baireddy2021, Spirkovska2010} thereby motivating this study.

The exponential growth of data production from modern instruments have made data-driven techniques and machine learning appealing to astronomers and telescope operators. However efforts in machine learning based anomaly detection are concentrated in scientific discovery rather than SHM, with approaches ranging from detecting unusual galaxy morphologies~\cite{Margalef-Bentabol2020,Storey-Fisher2021} to identifying new transients~\cite{Villar2021, Lochner2020, Ma2023}. Unfortunately, these techniques are not directly applicable to the multi-baseline autocorrelation-based spectrographic data obtained from radio observatories, due to increased data complexity, high dynamic range due to RFI, varying observation durations and frequency ranges as well as the feature compounding problem~\cite{Mesarcik2020b}. It must be noted this work makes use of up-stream data products in the form of spectrograms, that are produced by all radio telescopes thereby enabling its applicability to other instruments. 

The SHM anomaly detection problem differs from existing work for several reasons. Firstly, the data inspection performed by telescope operators involves analysing both known and unknown anomalies; where known anomalies should be classified into their respective classes and unknown anomalies should be differentiated from all other existing classes. This is in contrast with typical anomaly detection which is normally posed as one-class-classification problem. Furthermore, we find that class imbalance not only exists between the normal and anomalous classes (which is common for anomaly detection), but there is also strong imbalance between the anomalous classes. For these reasons, we propose a new framework for detecting and classifying SHM-based anomalies, that is capable of distinguishing both regularly occurring and rare events.    

Fundamentally, all anomaly detection approaches rely on learning representations of normal data and then measuring some difference between the learnt representations of normal and anomalous data~\cite{Chandola2009}. Recent developments in machine learning leverage pre-trained networks by fine-tuning them on specific classes of anomaly detection datasets~\cite{Roth2021, reiss2021, Tack2020}. However we show that it is not possible to directly apply these pretrained networks to astronomical data due to large differences compared to the natural images used for pretraining since these spectrograms are in the time-frequency domain. This being said, there has been effort made in pretraining paradigms for astronomical data~\cite{Hayat2021,Walmsley2022}, however similarly with anomaly detection applied to astronomy, these methods are implemented with imaged galaxy data and not the dynamic spectra necessary for SHM. For this reason we propose a new self-supervised learning paradigm that combines context prediction and reconstruction error~\cite{Doersch2015} as a learning objective and show that it is effective in learning robust representations of non-anomalous  time-frequency data. 

In this work we make the following contributions: (1) a new dataset consisting of 7050 manually labelled autocorrelation based spectrograms consisting of 10 different feature classes; (2) a generic self-supervised learning (SSL) framework that is effective in learning representations of time-frequency data with a high-dynamic range; (3)  a generic anomaly detection framework that can classify both commonly occurring known anomalies and detect unknown anomalies with a high precision, and (4) we show that our implementation achieves real-time performance for LOFAR.

We begin in the remaining part of the paper with an analysis of existing literature concerning anomaly detection in astronomy in Section~\ref{sec:related_work}, and Section~\ref{sec:dataset} documents our data selection strategy and outline the labelling process used for evaluation of this work. In Section~\ref{sec:method} we show the proposed SSL and anomaly detection frameworks. Finally, our results and conclusions are documented in Sections~\ref{sec:experiments} and \ref{sec:conclusions}.
    \section{Related work}
\label{sec:related_work}
Recent works that apply machine learning-based anomaly detection to astronomy have so far focused on only scientific discovery, using galaxy images, transient signals or light curves. In this work we apply machine learning-based anomaly detection to autocorrelation-based spectrograms obtained from the LOFAR telescope. This section unpacks the current landscape of machine learning-based anomaly detection and the recent developments in applying it to astronomy-related fields.    

\subsection{Machine learning-based anomaly detection}
Machine learning-based anomaly detection relies on modelling normal data and then classifying abnormality by using a discriminative distance measure between the normal training data and anomalous samples~\cite{Chandola2009}. Autoencoding models are a popular approach for learning latent distributions of normal data~\cite{Bergmann2019a, Bergmann2019, Pidhorskyi2018, An2015}. Anomaly detection using autoencoders can be performed either in the latent space using techniques such as One-Class Support Vector Machines (OC-SVM)~\cite{Scholkopf1999}, K-Nearest-Neighbours (KNN)~\cite{Bergman2020} or Isolation Forest (IF)~\cite{TonyLiu2008} or using the reconstruction error~\cite{Mesarcik2021}. The use of pretrained networks to obtain latent representations of normal data have also been successful in anomaly detection~\cite{Bergman2020,reiss2021,Roth2021}. By first training these models on an objective such as ImageNet classification~\cite{Fei-Fei2010}, they are able to generalise to other tasks such as anomaly detection.  Additionally, Self-Supervised Learning (SSL) has been shown to be invaluable for finding meaning representations of normal data~\cite{Yi2020, Li2021, Tack2020}. Here pretext tasks, which allow the model to learn useful feature representations or model weights that can then be used for other (downstream) tasks, are defined as learning objectives on the normal data such that the model can be fine-tuned for the downstream task of anomaly detection.  In both the SSL and pretrained cases, KNN-based measures can be used to distinguish anomalous samples from the normal training data~\cite{Bergman2020,Yi2020}.

In most machine learning-based anomaly detection, performance is evaluated according to the Single-Inlier-Multiple-Outlier (SIMO) or Multiple-Inlier-Single-Outlier (MISO)~\cite{Burlina2019} settings on natural image datasets such as MVTecAD-\cite{Bergmann2019a}. With this paradigm in mind, we find that anomaly detection in the radio astronomical context is a Multiple-Inlier-Multiple-Outlier (MIMO) problem. In effect, anomaly detection formulations that make a strong assumption about the number of inliers or outliers are not directly applicable to the radio observatory setting due to the increased problem complexity. Furthermore, we find methods that rely on pretraining with natural images to be ill-suited to the spectrograms used in this work, due to differences in dynamic range and SNR as shown by \cite{Mesarcik2022}. 

Efforts have been made for detecting anomalies in light-curves and spectra in works such as Astronomaly~\cite{Lochner2020} and transients in~\cite{Malanchev2021}. Astronomaly is an active learning framework developed for the classification of unusual events in imaged data or light curves at observatories to aid with scientific discovery. This being said, it follows closely with generic anomaly detection methods, where normal data is first projected to a latent representation and metrics such as IF are used to distinguish normal training samples from anomalous testing samples at inference time. Although Astronomaly assumes a MIMO context, it is still only able to detect unknown anomalies (or at least says all anomalies belong the same class). This is in contrast with our work, where \texttt{ROAD} is capable of both distinguishing all known anomaly classes with a high precision as well as detecting unknown or rare anomalies. 

Deep generative neural networks are also used for anomaly detection. Works by \cite{Villar2021,Mesarcik2020b,Ma2023} have shown that the Variational Autoencoders (VAEs) can be used for anomaly detection with astronomical data. Whereas \cite{Margalef-Bentabol2020, Storey-Fisher2021} show that Generative Adversarial Networks (GANs) are effective in learning representations of normal images of galaxies thereby enabling reconstruction-error based anomaly detection. In work by \cite{Zhang2019} GANs have also been shown to be effective in the Search for Extraterrestrial Intelligence (SETI) anomaly detection context. However, we find that our SSL method is more stable during training and better suited at anomaly detection using time-frequency data, that has a high dynamic range, $\in [1,100]$~dB, and a low SNR for cross polarised features in the \textit{xy} and \textit{yx} stokes parameters such as the galactic plane. 

\subsection{Representation learning in astronomy}
As already mentioned, learning representations of high-dimensional data is essential to the anomaly detection problem. For this reason, among many others, tremendous effort has been made to find methods that learn robust projections of high-dimensional data~\cite{He2022, Chen2020,Grill2020,Doersch2015}. These successes have materialised in the astronomical community with results mostly in the galaxy classification domain. \cite{Walmsley2022} show that by pretraining on the Galaxy Zoo DECaLS~\cite{Walmsley2021} dramatically improves model performance for several downstream tasks. Furthermore, \cite{Hayat2021} show how contrastive learning can be applied to galaxy photometry from the Sloan Digital Sky Survey (SDSS)~\cite{Gunn1988}. The authors show that with novel data augmentations, they can achieve state of the art results on several downstream tasks. Furthermore, several additions and modifications have been made to the reconstruction error-based loss functions of autoencoders. \cite{Mesarcik2020b} show how using both magnitude and phase information in VAEs improves performance of finding representations of astronomical data,  whereas~\cite{Villar2021} use a recurrent adaption of a VAE to make training more suitable to light-curve data. Similarly, \cite{Melchior2022} show how the inclusion of self-attention mechanisms and redshift-priors into the latent projection of autoencoders, can improve the learnt representations of galaxy spectra. 

In this work, we demonstrate that using a simple adaption of a context-prediction self-supervised loss~\cite{Doersch2015} we are effective in learning robust representations of spectrograms from the LOFAR telescope. Our Radio Observatory Anomaly
Detector (ROAD) outperforms existing autoencoding models by a large margin on anomaly detection benchmarks. 

\subsection{Real-time scientific data processing}
To cope with the increasing data-rates from modern scientific instruments~\cite{Norris2010,VanHaarlem2013,Nan2011,LaPlante2021} real-time algorithms have been developed for scientific data pipelines. Real-time methods for RFI detection~\cite{Sclocco2016, Sclocco2019, Morello2021}, calibration~\cite{Prasad2014}, Fast Radio Burst (FRB) detection~\cite{Connor2018} and correlation~\cite{vanNiewpoort2011, Romein2010} have been essential to modern radio telescope operations. However, very few machine learning techniques have been shown to be effective in real-time. In seminal work by~\cite{George2018} machine learning gravitational wave detection algorithms were implemented in real time. Furthermore~\cite{Muthukrishna2022} show that Temporal Convolutional Networks (TCNs) can be implemented to detect transient anomalies in real time. To demonstrate the effectiveness of our work in the context of radio observatories we investigate the computation performance and throughput of the proposed system. We show that our system is real-time in the context of the LOFAR telescope data processing pipeline. 
    \section{Dataset}
\label{sec:dataset}
\begin{table*}[h!]
    \centering
   \resizebox{\linewidth}{!}{
    \begin{tabular}{|l|l|l|l|l|l|}
        \hline
        \textbf{Category} & \textbf{Description} & \textbf{Band} & \textbf{Polarisation} &\textbf{Occurrence rate} & \textbf{ \# Samples}\\
        \hline
        \hline
         Normal & All non-characterised effects & Both & All & - & 4687 \\
        \hline
        \hline
        \textbf{Data processing}& & & & &\\
        \hline
         First order data loss & Data loss from consecutive time and/or frequency channels & Both & All & 0.02 & 146 \\
        \hline
         Second order data loss & Data loss from single frequency and/or single time channels & Both & All & 0.04 & 283 \\
        \hline
        \hline
        \textbf{Electronic systems}& & & & &\\
        \hline
         High noise element & High power disturbances caused by miscellaneous events& Both & All & 0.01 & 88 \\
        \hline
         Oscillating tile & Amplifier going into oscillation & High & All & 0.01  & 56 \\
        \hline
        \hline
        \textbf{Astronomical events}& & & & &\\
        \hline
        Source in side-lobes & A-team source passing through side-lobes  & High & All & 0.06 & 446\\
        \hline
         Galactic plane & Galactic plane passing through the main lobe of the antenna & Both & Cross  & 0.08 & 550 \\
        \hline
         Solar storm & Strong emissions from the sun  & Low & All & 0.02 & 147 \\
        \hline
        \hline
        \textbf{Environmental effects}& & & & &\\
        \hline
        Lightning & Lightning storm & Both & All & 0.06 & 389 \\
        \hline
         Ionospheric RFI reflections & RFI reflected from the ionosphere  & Low & All & 0.04 & 261 \\ 
         \hline
    \end{tabular}}
    \caption{Categorisation of data processing, electronic, astronomical and environmental anomalies in the \ROAD~dataset. Where A-team sources refer to the four brightest persistent radio sources in the northern sky.}
    \label{tab:dataset}
\end{table*}

We created a new dataset for anomaly detection in radio observatories and document the data selection, preprocessing and labelling strategy used in this section. Applying machine learning to radio astronomical datasets poses a significant challenge, particularly when using time-frequency data. Methods for preprocessing and data selection need to be carefully considered, due to issues such as high-dynamic range (due to RFI among other events), combining thousands of baselines for a single observation, having complex-valued data with multiple polarisations, feature compounding and many more. An additional challenge with applying machine learning to radio astronomy is the lack of labelled time-frequency datasets from radio telescopes as well as the availability of expert knowledge and the cost associated with creating a dataset. 

\subsection{Observation selection and preprocessing}
\label{sec:processing}
The~\ROAD~dataset is made up of observations from the Low Frequency Array (LOFAR) telescope ~\cite{VanHaarlem2013}. LOFAR is comprised of 52 stations across Europe, where each station is an array of 96 dual polarisation low-band antennas (LBA) in the range 10–90 MHz and 48 or 96 dual polarisation high band antenna antennas (HBA) in the range 110–250 MHz. The signals received by each antenna are coherently added in the station beamformer, resulting in each sub-band being approximately 200~kHz wide. These signals are then transported to the central processor to be correlated with a minimum channel width of about 0.7~kHz.  This data product is referred to as a visibility and is the data representation used in this work. In contrast, other radio astronomical use-cases where machine learning-based anomaly detection has been applied (such as detecting unusual galaxy morphologies) use an additional calibration step as well as a 2D Fourier transform and gridding to obtain sky maps. 

The visibility data is four dimensional with the dimensions corresponding to time, frequency, polarisation, and baseline. Different science cases result in different observing setups, which dictate the array configuration (i.e. the number of baselines used), the number of frequency channels ($N_f$), the time sampling as well as the overall integration time ($N_t$) of the observing session. Furthermore, the dual-polarisation of the antennas result in a correlation product ($N_\text{pol}$) of size 4. In the case of this work, we make use of only the autocorrelations produced by LOFAR. We do this to minimise the labelling overhead and data size, as well as to simplify the potential feature compounding problem~\cite{Mesarcik2020b}. 

As already mentioned, the required resolution of modern instruments cause the data products to be relatively large. The data size of an observation that consists only of autocorrelations is given by $N_\text{{auto}} = N_t N_f N_{st} N_\text{{pol}} N_{\text{bits}}$, where $N_{st}$ is number of stations. This means that a 10 hour observation with a 1 second integration time, a 1KHz channel resolution with a 50MHz bandwidth and a 32-bit resolution can result in observations sizes in the order of terabytes. As this is orders of magnitude larger than amount of memory available on modern GPUs that are used for training of machine learning algorithms, the data is sub-sampled in time and frequency according to~\cite{Mesarcik2020b} to result in observations in the order of 1 gigabytes. 

Deep learning architectures typically require equally-sized inputs, however LOFAR observations can have a varying number of time samples and/or frequency bands. Therefore, additional resizing of the intermediate visibilities is done by resizing all observations to $(256, 256)$ bins in time and frequency. This means that observations with fewer than 256 time samples are interpolated and those with more are down-sampled. Furthermore, as the autocorrelations contain no phase information, we use only the magnitude component of each spectrogram. 

It must be noted that this processing does modify the morphologies of certain features, particularly those present with a low time resolution. However as this preprocessing step is consistent across all spectrograms, the overall effects on the anomaly detector and classifier are negligible. In future work, we plan to associate the labels with the full resolution LOFAR data from the Long Term Archive (LTA)\footnote{\href{https://lta.lofar.eu/}{https://lta.lofar.eu/}} and apply it to (256, 256) crops of the full resolution spectrograms.

We selected 111 observations from the LOFAR LTA comprising of a broad set of science use cases and the corresponding observing setups. Of the selected observations, we use the autocorrelations from 2436 LBA baselines and 4617 HBA baselines from an observation period between 2019 and 2022.  

\begin{figure*}[hbtp!]
     \centering
     \begin{subfigure}[b]{0.49\textwidth}
         \centering
         \includegraphics[trim={0.4cm 0.0cm 2.0cm 0cm},clip,width=0.8\textwidth]{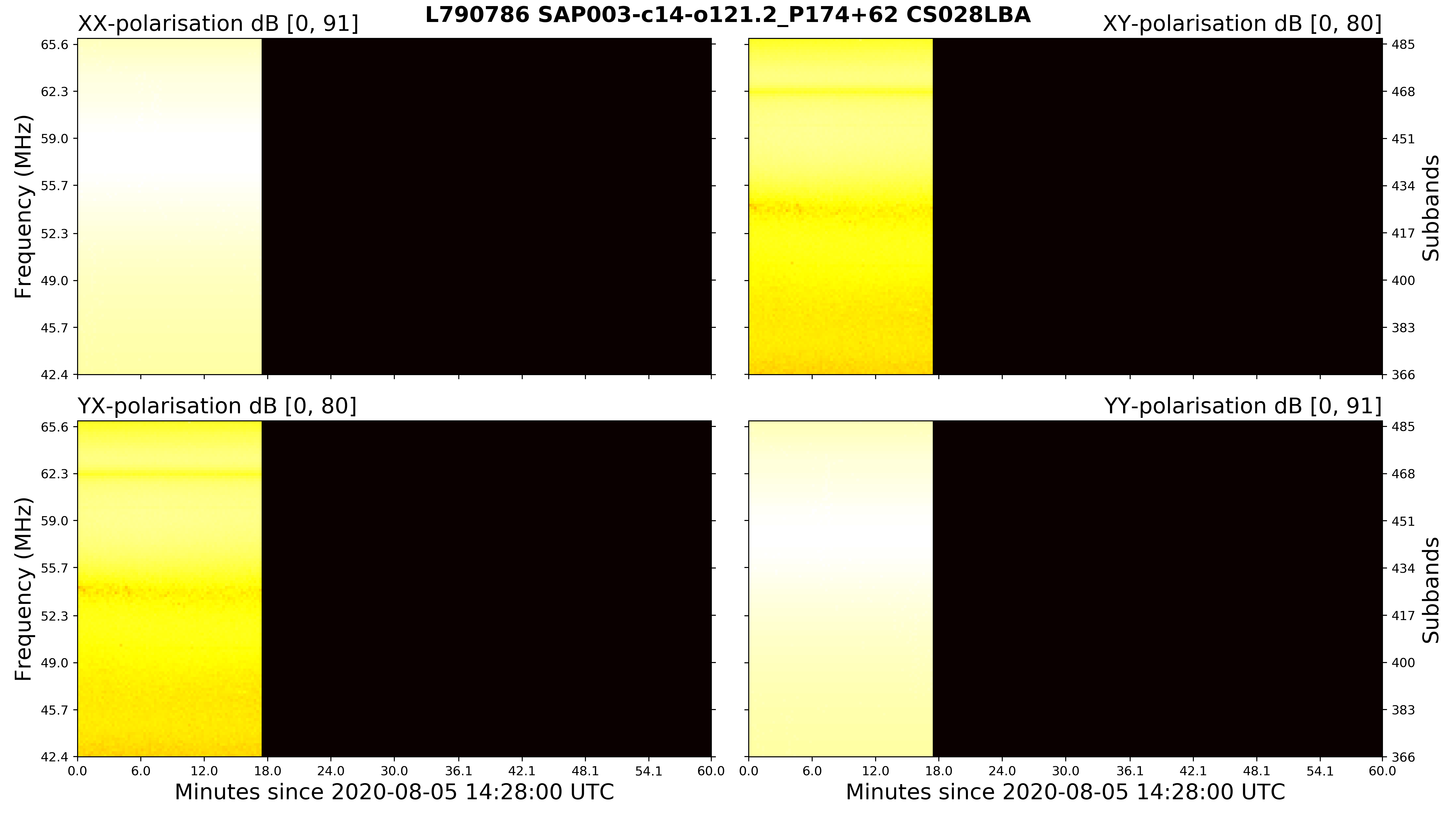}
         \caption{First order data loss}
         \label{fig:first_order_data_loss}
     \end{subfigure}
     \begin{subfigure}[b]{0.49\textwidth}
         \centering
         \includegraphics[trim={0.4cm 0.0cm 2.0cm 0cm},clip,width=0.8\textwidth]{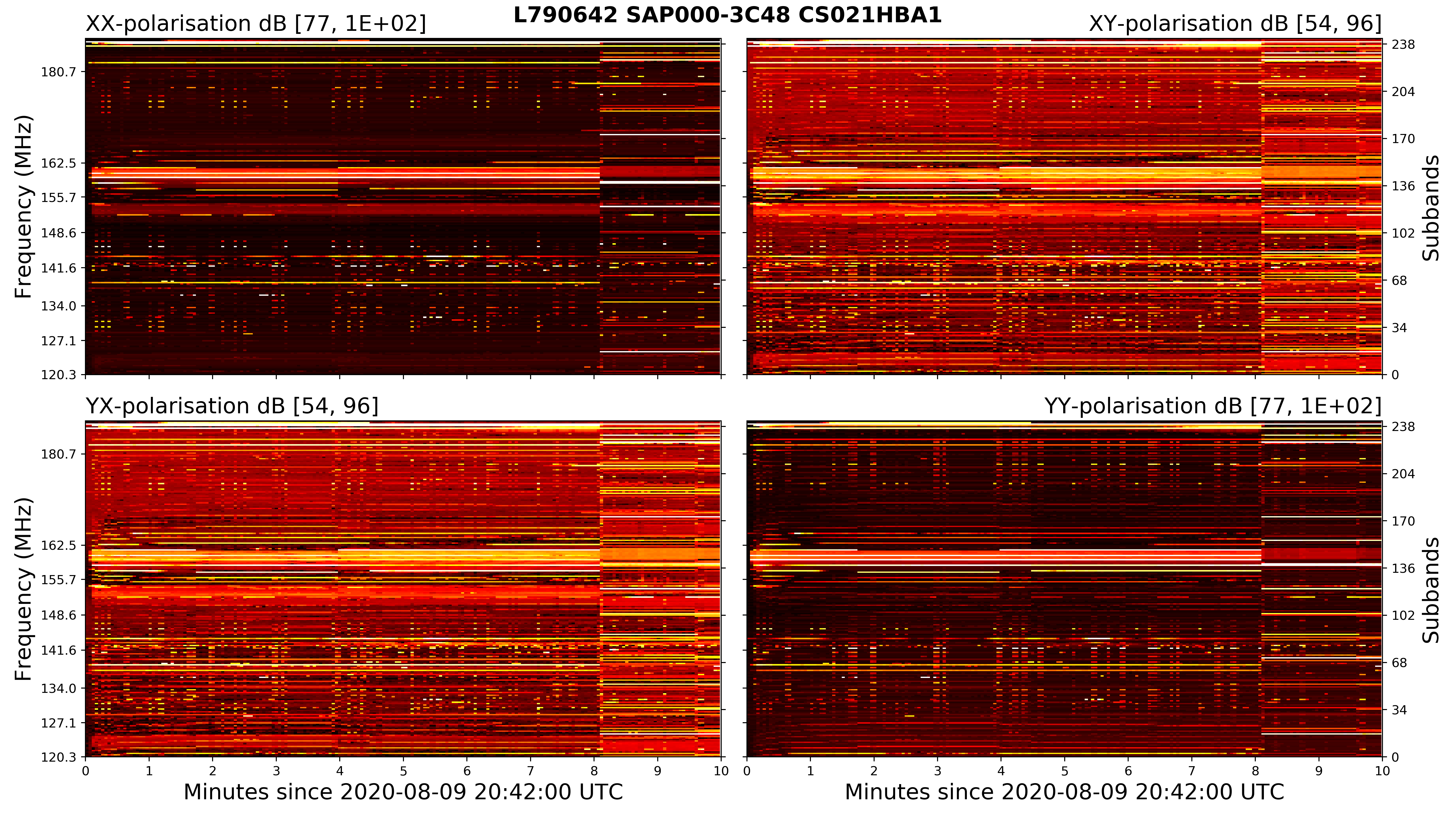}
         \caption{Oscillating amplifier}
         \label{fig:oscilating_tile}
     \end{subfigure}
     ~
     \begin{subfigure}[b]{0.49\textwidth}
         \centering
         \includegraphics[trim={0.4cm 0.0cm 2.0cm 0cm},clip,width=0.8\textwidth]{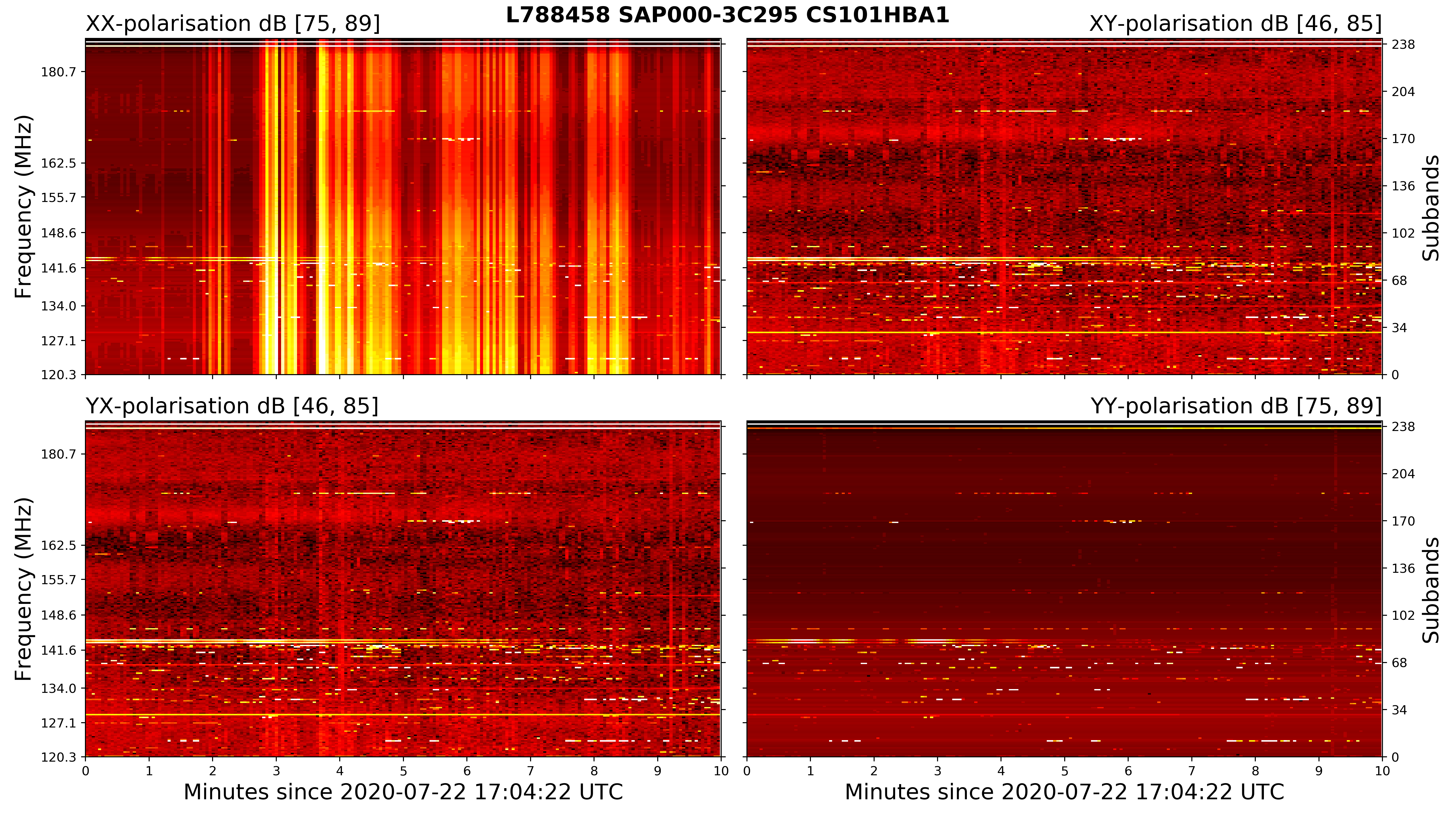}
         \caption{High noise element}
         \label{fig:high_noise}
     \end{subfigure}
     \begin{subfigure}[b]{0.49\textwidth}
         \centering
         \includegraphics[trim={0.4cm 0.0cm 2.0cm 0cm},clip,width=0.8\textwidth]{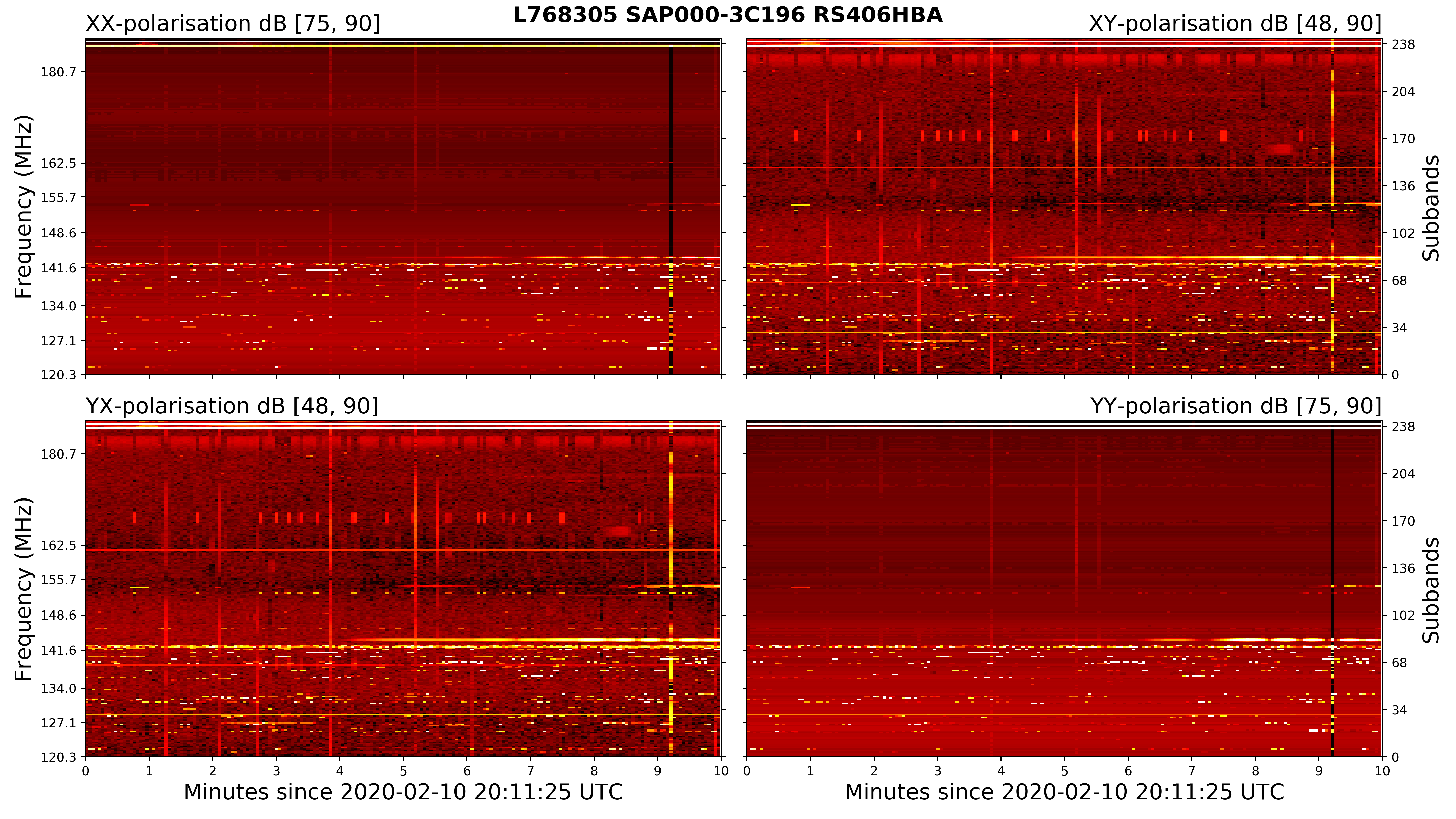}
         \caption{Lightning storm}
         \label{fig:lightning}
     \end{subfigure}
     ~
     \begin{subfigure}[b]{0.49\textwidth}
         \centering
         \includegraphics[trim={0.4cm 0.0cm 2.0cm 0cm},clip,width=0.8\textwidth]{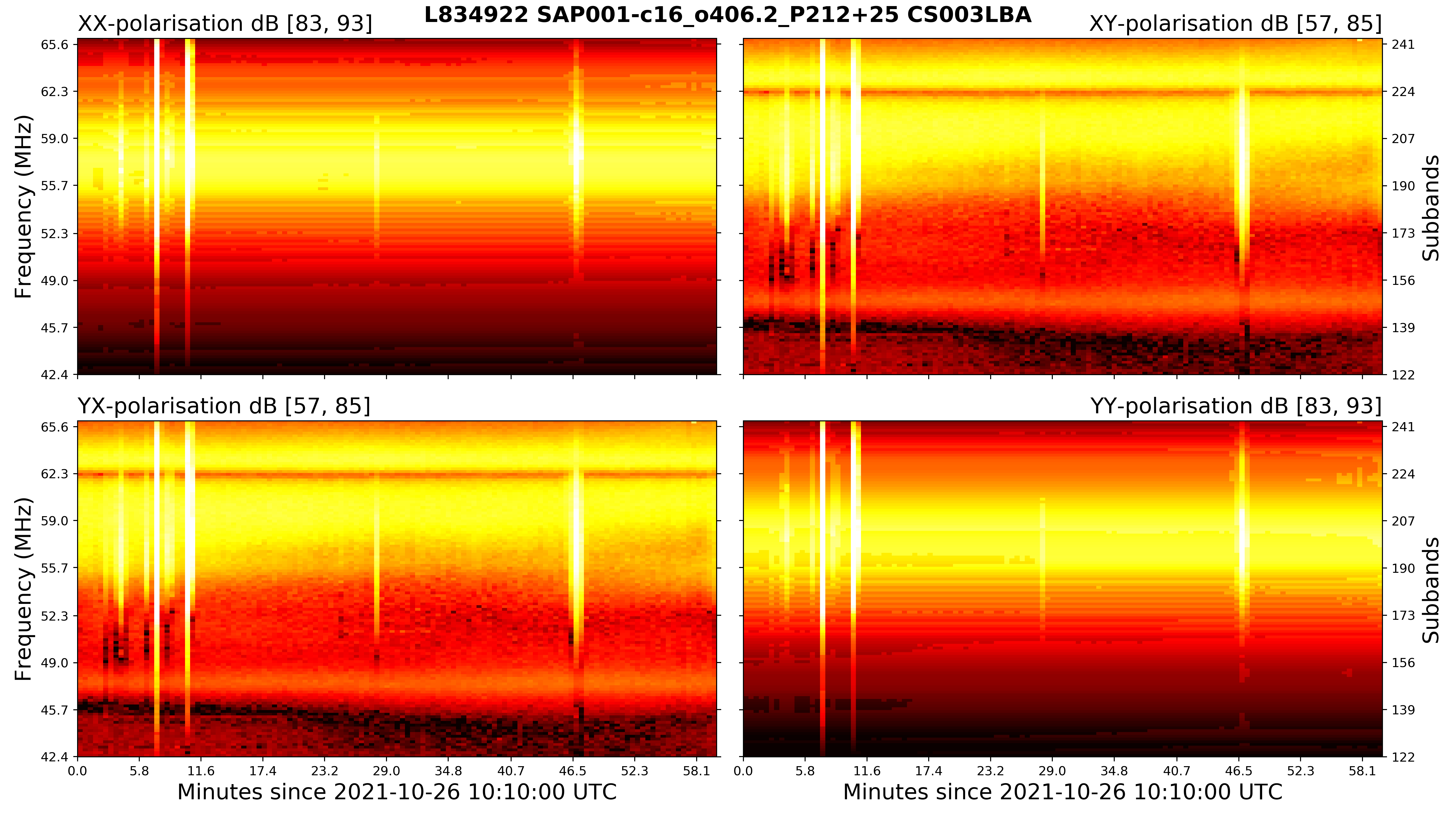}
         \caption{Solar storm with ionospheric effects}
         \label{fig:solar}
     \end{subfigure}
     \begin{subfigure}[b]{0.49\textwidth}
         \centering
         \includegraphics[trim={0.4cm 0.0cm 2.0cm 0cm},clip,width=0.8\textwidth]{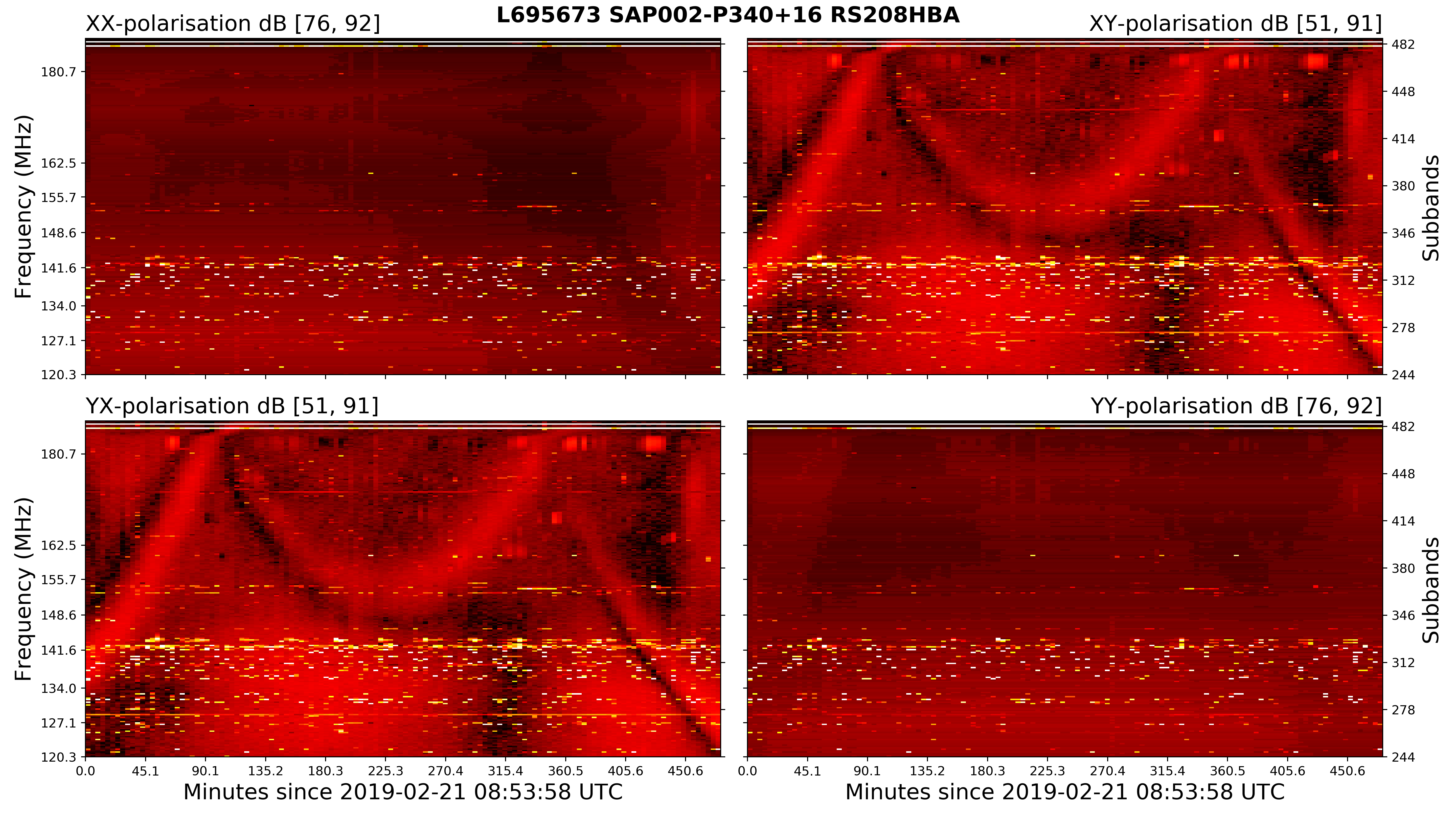}
         \caption{Galactic plane}
         \label{fig:source_in_sidelobes_1}
     \end{subfigure}
    ~
     \begin{subfigure}[b]{0.49\textwidth}
         \centering
         \includegraphics[trim={0.4cm 0.0cm 2.0cm 0cm},clip,width=0.8\textwidth]{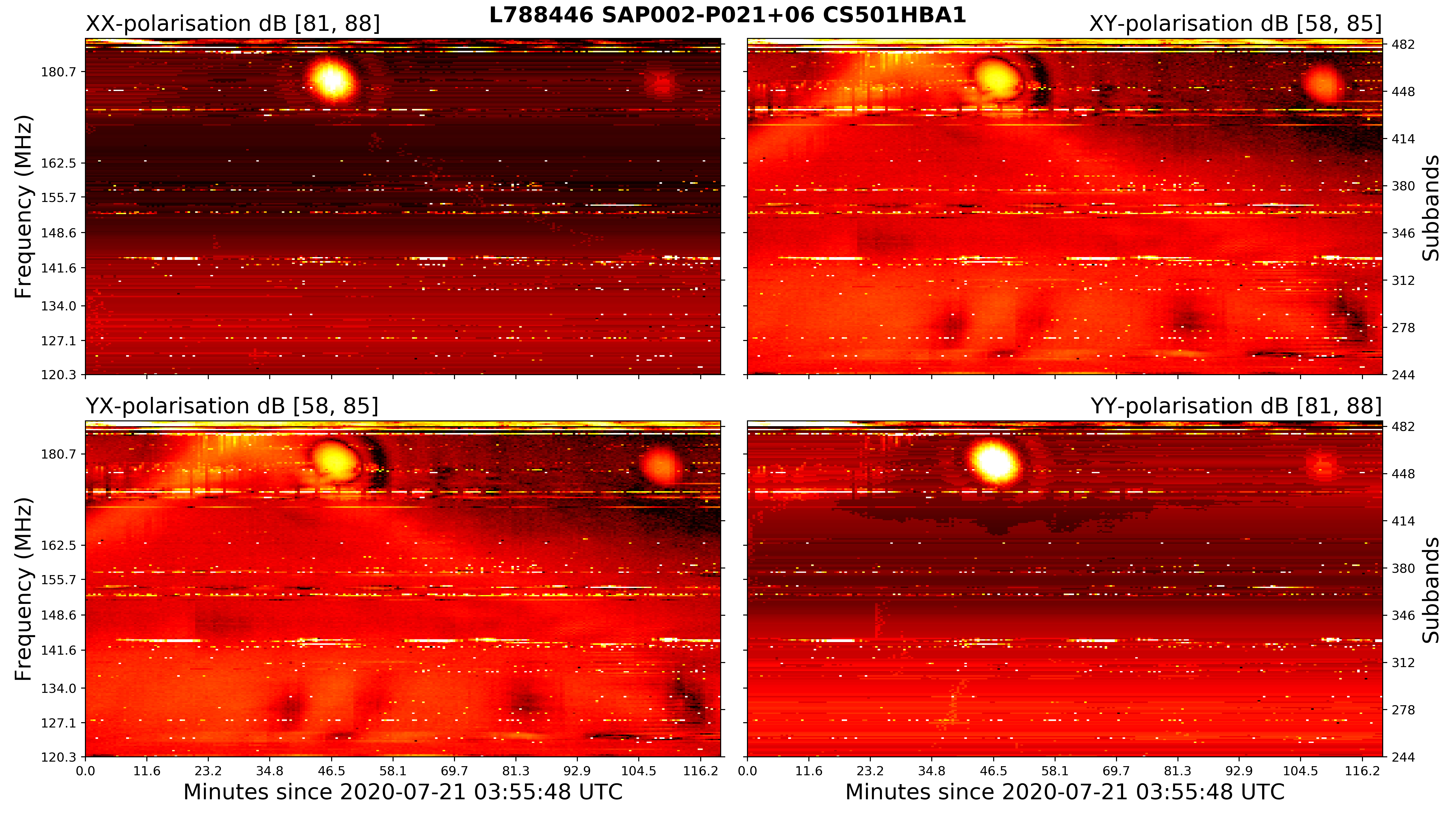}
         \caption{Source in the antenna side-lobes}
         \label{fig:source_in_sidelobes}
     \end{subfigure}
     \begin{subfigure}[b]{0.49\textwidth}
         \centering
         \includegraphics[trim={0.4cm 0.0cm 2.0cm 0cm},clip,width=0.8\textwidth]{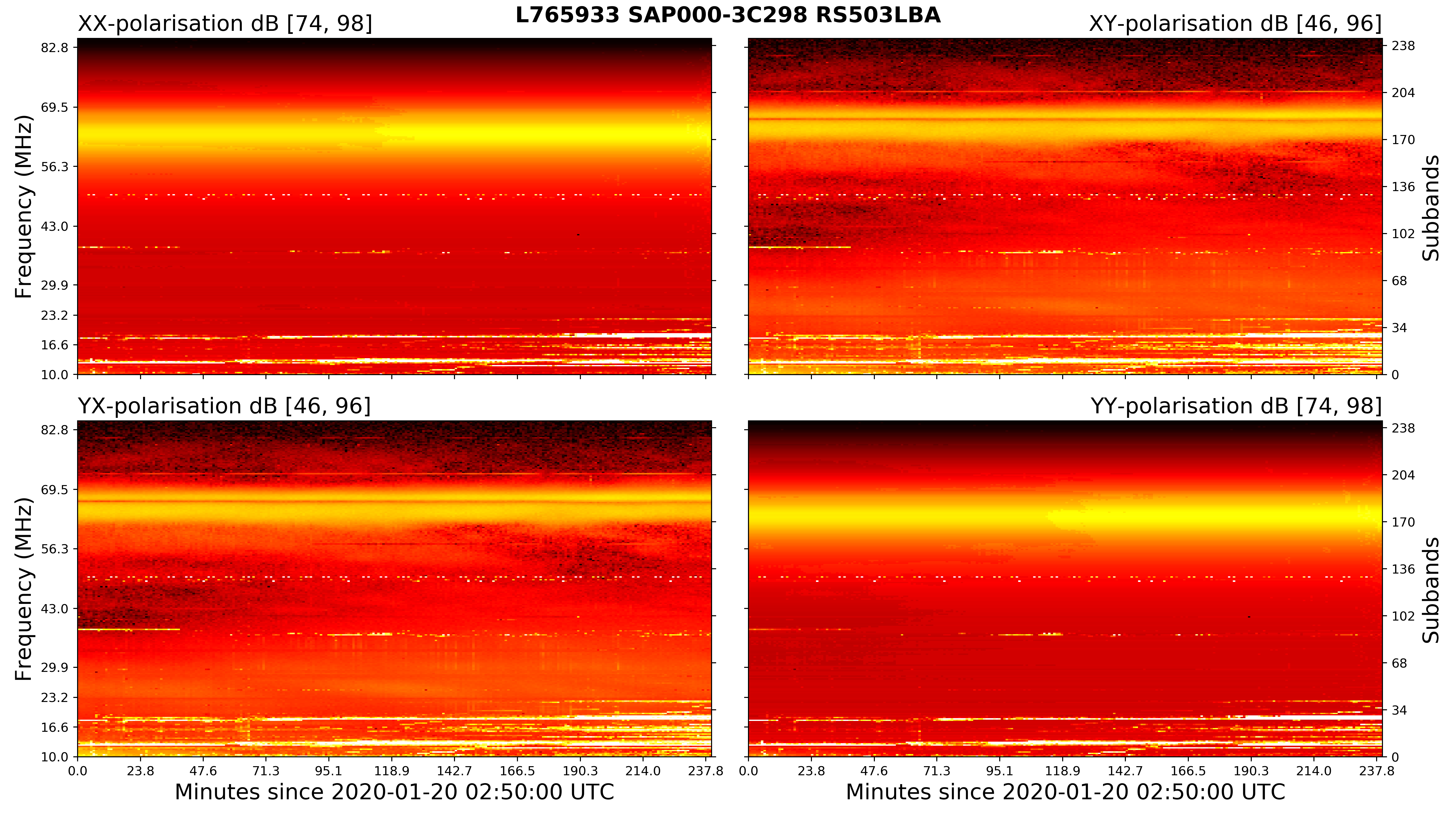}
         \caption{Low frequency ionospheric RFI reflection}
         \label{fig:ionosphere}
     \end{subfigure}
      ~
     \begin{subfigure}[b]{0.49\textwidth}
         \centering
         \includegraphics[trim={0.4cm 0.0cm 2.0cm 0cm},clip,width=0.8\textwidth]{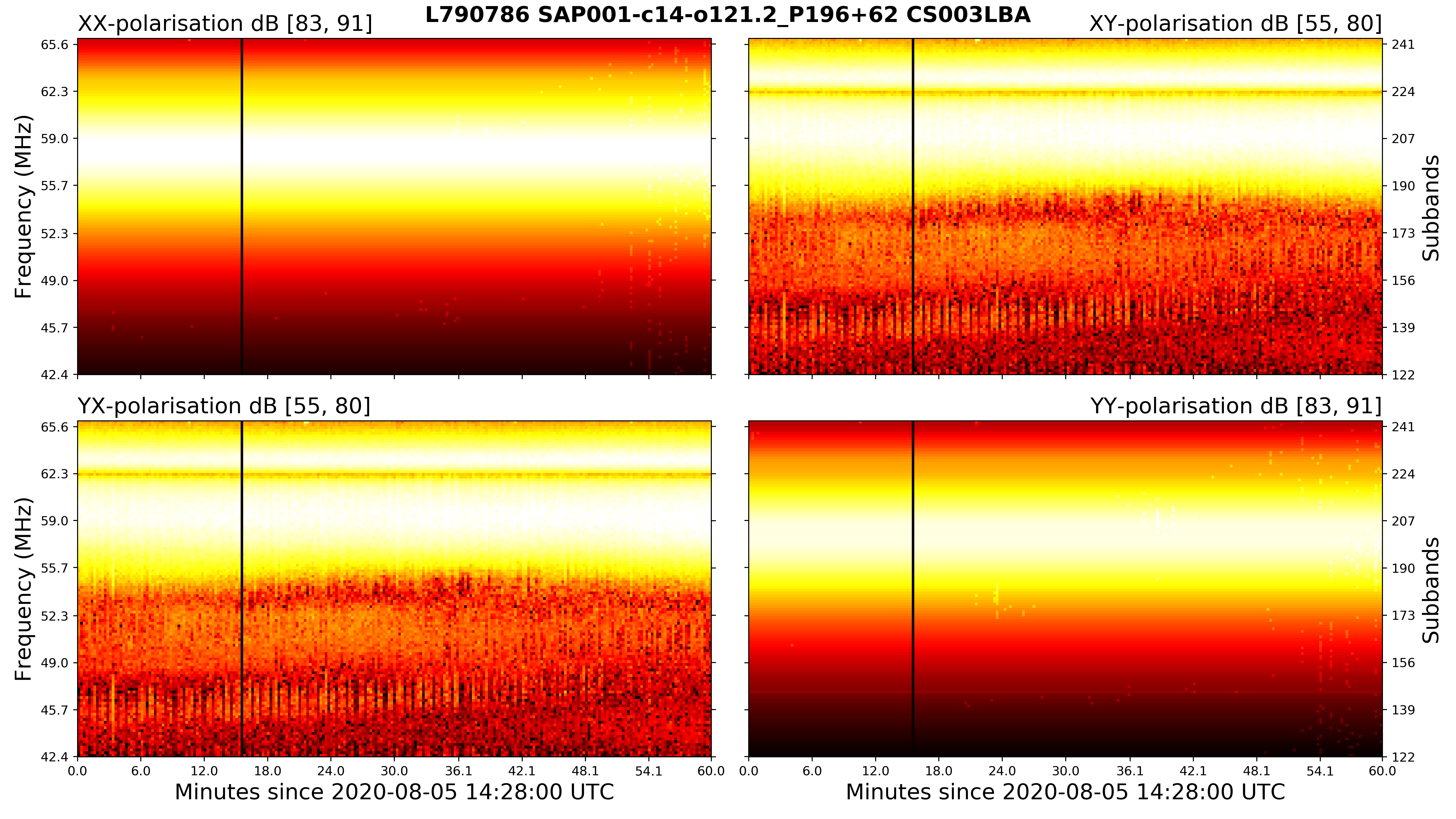}
         \caption{Second order data loss}
         \label{fig:second_order_data_loss}
     \end{subfigure}
     \begin{subfigure}[b]{0.49\textwidth}
         \centering
         \includegraphics[trim={0.4cm 0.0cm 2.0cm 0cm},clip,width=0.8\textwidth]{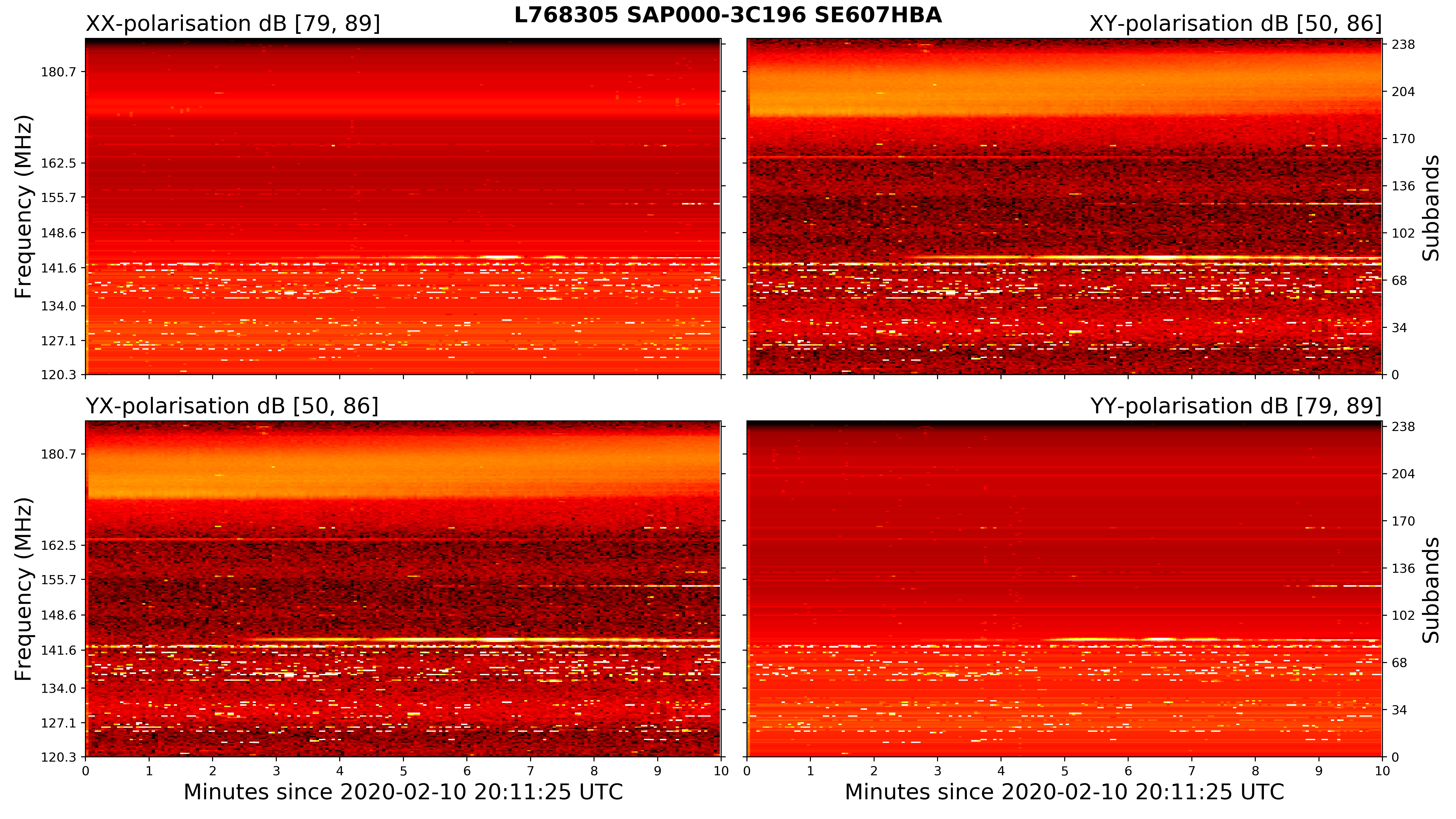}
         \caption{Normal}
         \label{fig:normal}
     \end{subfigure}
     \caption{Illustration of 10 examples from the \ROAD~dataset.}
     \label{fig:dataset}
\end{figure*}

\subsection{Labelling methodology}
The \ROAD~dataset contains 10 classes which describe various system-wide phenomena and anomalies from data obtained by the LOFAR telescope. These classes are categorised into 4 groups: data processing system failures, electronic anomalies, environmental effects, and unwanted astronomical events.  Table~\ref{tab:dataset} shows the classes used as well as the description of events, their band and polarisation in which they occur. We note that the term \textit{anomaly} is used liberally in this context, while low power effects (that are only present in the cross polarisations) such as the galactic plane passing through an observation are somewhat unavoidable. Nonetheless, for observations with extremely low SNR such as The Epoch of Re-ionisation of the Universe (EoR)~\cite{Yatawatta2013}, the galactic foreground signals need to be identified and removed. For this reason, we include such events in the \ROAD~dataset. Furthermore, we do not consider classes which track the systematic corruptions caused by ionospheric disturbances. This is because the ROAD dataset was created using data from the period of the minimum of the past Solar cycle. Thus the statistics for corruption effects like scintillation are poorly represented in high-band and low-band data (although low-band data tracks better these events due to the frequency dependence of the signals). In future work, we plan to extend the dataset to consist of classes relating to more ionospheric disturbances.

Our labelling approach took into consideration anomalies which occurred at both the station- and observation-levels. For example, events such as lightning storms and high-noise events can look fairly similar, especially in the down-sampled context. However, lightning storms are geographically-bound to affect all stations in a certain region therefore only occurring at the station-level. Additionally, lightning is highly correlated across stations in time, with minimal delay between the recorded events in each station. Whereas high-noise events usually affect only a single antenna at a time with no time dependency between antennas and stations.  By this logic, all stations bound to the same geographic location with broadband high power events across all polarisations that are correlated in time were considered to be corrupted by lightning storms, whereas individually affected stations were labelled as high-noise events.  

We make a distinction between first and second order events; for example the first order data-loss event corresponds to dropped information from consecutive time samples and/or frequency bands and second order is for a single time sample/frequency band. We find this a useful distinction as the root cause of these events is different. In the case of first order data loss events, the problem can be traced to the correlator pipeline, whereas the second order events are most-likely from to type conversion overflows due to strong RFI. Additionally, we note some overlap between class labels, for example it is common for a high power noise events to trigger instability in an amplifier causing it to oscillate, however the precise point of transition is often hard to distinguish these events from each other. 

We labelled the dataset using LOFAR observations that were down-sampled and preprocessed as described in Section~\ref{sec:processing}. We made multiple train-test splits during experimentation to ensure consistent performance across models. Furthermore, the \ROAD~dataset is publicly available \footnote{\href{https://zenodo.org/record/8028045}{https://zenodo.org/record/8028045}}. The file is in the \texttt{hdf5} format and consists of fields corresponding to the raw data, labels, frequency bands information, station name and source observation. Figure~\ref{fig:dataset} illustrates 8 of the 10 classes labelled in the available dataset for brevity. 

\subsection{Class imbalance}
\label{sec:subsample}
Due to the nature of anomaly detection, the number of normal samples greatly outnumbers anomalous ones. In the case of the \ROAD~dataset, and the LOFAR telescope more generally, we find there is not only a class imbalance between normal and anomalous classes but also among the anomalous classes. For example, commonly occurring astronomical signals, such as the galactic plane, are far better represented in the observations than unlikely events like the amplifiers oscillating. Practically, this means that when we separate the samples into testing and training sets we need to also maintain the same occurrence rates with respect to the rates in the original dataset. 

In effect, we down-sample the testing data such that the occurrence rate (shown in the second last column of Table~\ref{tab:dataset}) is maintained for evaluation. This means that each model needs to be tested multiple times with new samples taken from the testing pool of anomalous samples to effectively evaluate its performance. We evaluate each model 10 times with different random seeds to ensure accurate reporting.
    \section{The radio observatory anomaly detector}
\label{sec:method}
As outlined in preceding sections \ROAD~is designed to detect both previously unseen system behaviours as well as classify known-anomalies observed by the LOFAR telescope. To accommodate these requirements, we find it necessary to combine two approaches; supervised classification, as well as self-supervised anomaly detection. This section outlines the motivations and design decisions made for the implementation of \ROAD. 

\subsection{Problem formulation}
Given the $i^{\text{th}}$ spectrogram $V_i(\nu, \tau, b, p)$ from the dataset and model $m$ with parameters $\theta_m$;  we would like to predict whether an anomaly is present and which class it belongs to, if it is a known event, such that, 

\begin{equation}
    m_{{\theta}_m}(V_i(\nu, \tau, b, p))=
    \begin{cases}
        0 &,\text{if normal}\\
        [1,N] &,\text{if known anomaly}\\
        N+1 &,\text{if unknown anomaly}\\
     \end{cases}
\label{eq:problem}
\end{equation}

where $\nu$, $\tau$, $b$ and $p$ are the indexes corresponding to frequency band, time sample, baseline and polarisation, respectively and N is the number of known anomaly classes. Supervised approaches assume that each class is represented in the training set and try to minimise the following loss function

\begin{equation}
    \mathcal{L}_{\text{sup}} = \min_{\theta_m} \sum_i \mathcal{H}(m_{\theta_m}(V_i(\nu, \tau, b, p), l_i)
\end{equation}

\begin{figure}[h!]
    \centering
    \includegraphics[trim={4.3cm 3.7cm 6.75cm 3.3cm},clip, width=\linewidth]{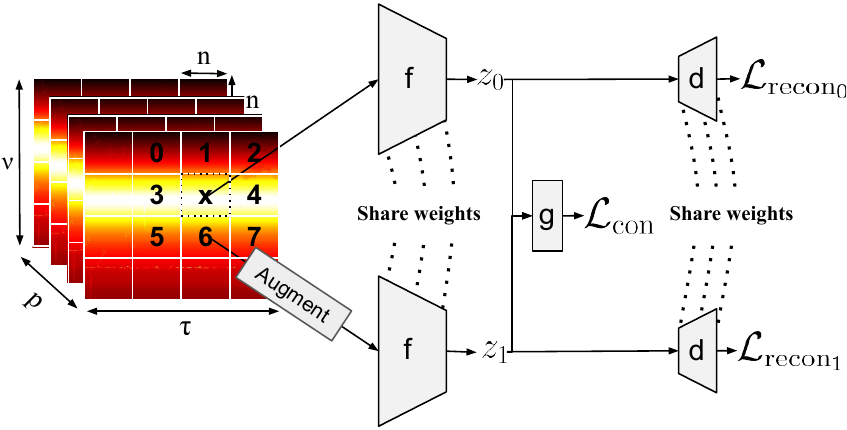}
    \caption{Illustration of the self-supervised training procedure used in \texttt{ROAD}, we use random-cropping for augmentation.}
    \label{fig:ssl_training}
\end{figure}
\begin{figure*}
    \centering
     \resizebox{\linewidth}{!}{
    \includegraphics[trim={1.5cm 4.5cm 3.5cm 4.5cm},clip, width=\linewidth]{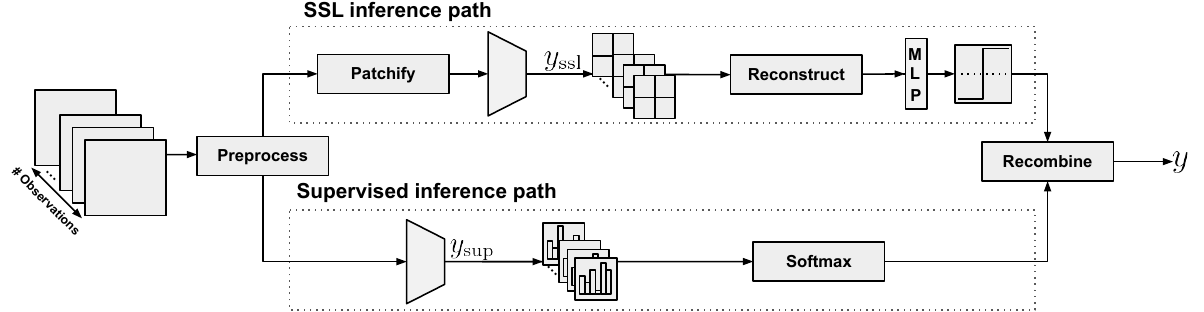}}
    \caption{Illustration the inference pipeline of ROAD; where we combine both supervised and self supervised learning to effectively detect radio observatory based anomalies.}
    \label{fig:road_diagram}
\end{figure*}


where $\mathcal{H}$ is an entropy-based measure of similarity and $l$ is the encoded vector of labels corresponding to the contents of $V$. During inference, the supervised classifier produces estimate of which classes are most probable in a given spectrogram, and the \textit{argmax} function selects the most likely classification as shown in the bottom half of Figure~\ref{fig:road_diagram}. However, as illustrated by the results in Section~\ref{sec:experiments}, the performance of such a supervised classifier severely deteriorates when exposed to unseen or out of distribution (OOD) classes during testing. To remedy this, we disentangle the two model objectives, namely we use a supervised classifier to identify the known classes present in the training set and a self-supervised anomaly detector to classify unseen anomalies.  
\subsection{Self supervised representation learning}
Self Supervised Learning (SSL) methods learn useful feature representations by training on secondary objectives called \textit{pretext tasks}, so that once trained, the model weights can be utilised for downstream applications. We define two pretext tasks that allow the model to learn useful representations for anomaly detection in astronomical data: context-prediction and reconstruction error. Context-prediction is a pretext task that makes a model classify the positional relationship between two patches taken from the same image. The two patches are projected to some latent representations $z_0$ and $z_1$ using a backbone network $f$, while keeping track of their position label, $c$, on a $3\times3$ grid as proposed by~\cite{Doersch2015}. Then, using $g$, a 2-layer Multi-layer Perceptron (MLP), we classify the positional relationship from the latent representations; as given by 

\begin{equation}
    \mathcal{L}_{\text{con}} = \sum_i \sum_j \mathcal{H}(g(z_{i,j,0}, z_{i,j,1}),c_j)
\end{equation}

where $i$ corresponds the the index of each spectrogram, $j$ is the index of each context-pr edition pair in a single spectrogram and $c_j$ is the positional label. Additionally, to ensure the model does not learn positional relationships based purely on the bordering values of each patch, we augment each neighbour in the training process. In the implementation, we randomly crop the patches between 100\% and 75\% of their original size followed by resizing them to their original dimensions. We illustrate the context prediction loss and patch selection in Figure~\ref{fig:ssl_training}.

Furthermore, to enforce consistency across the representations of similar looking patches we use reconstruction error. Reconstruction error maintains consistency by ensuring that two patches which share common features in visibility space, should occupy nearby locations in the latent space and therefore should be reconstructed similarly. The reconstruction loss is given by

\begin{equation}
    \mathcal{L}_\text{recon} = \sum_i \sum_j |V_{i,j,0} - d(z_{i,j,0})| + |V_{i,j,1} - d(z_{i,j,1})|
\end{equation}

where $d$ is a de-convolutional decoder that should have significantly fewer parameters than the backbone network $f$. We do this to ensure that the model has more capacity to learn suitable representations instead of prioritising reconstruction. For completeness we represent the full SSL learning objective as 

\begin{equation}
    \mathcal{L}_\text{SSL} = \lambda \mathcal{L}_{\text{con}} + (1-\lambda) \mathcal{L}_{\text{recon}} + \lambda_{\text{reg}}\sum_i\sum_j(z_{i,j,0}^2 + z_{i,j,1}^2)
\end{equation}

where $\lambda$ is a hyper-parameter which changes the influence of each component of the loss. Additionally, we use regularisation in the form of minimising the square size of the latent projections $z$. Regularisation is used in order to enforce the most compact representations in $z$. We experimentally select $\lambda = 0.5 $ and $\lambda_{\text{reg}}=1\times10^{-6}$, and illustrate $\lambda$'s impact in Section~\ref{sec:experiments}.

\subsection{Distinguishing normal from anomalous samples}
Although we have described a method for learning representations of normal data, the model is incapable of accurately discriminating between normal and anomalous samples. Several options exist for anomaly detection when utilising the learnt representations of normal training data methods. The simplest involves measuring the distance between a given sample and the normal training data~\cite{Bergman2020} using a K-Nearest-Neighbour (KNN) lookup. This assumes that larger distances correspond to more anomalous samples. However as we already make use of some of the labelled data for the supervised classifier we find it beneficial to fine-tune a shallow MLP on top of SSL representations to perform anomaly detection. As the SSL-backbone learns representations on the patch-level and \ROAD~dataset labels are on the spectrogram-level, we first need to concatenate the latent representations of each patch to return to the correct dimensionality before training the MLP. Notably, we propagate the gradients during fine tuning through both the MLP and the backbone network, $f$, such that the distance between normal and anomalous representations at the spectrogram-level are consolidated. We show in Section~\ref{sec:experiments} that fine-tuning dramatically outperforms random initialisation and KNN-based anomaly detection. Furthermore, we find that using the fine-tuned approach dramatically improves the time-complexity of the system. 

Additionally, we need to determine how to threshold the anomaly scores produced by either the fine-tuned models or the KNN-distance based approach. Here we utilise the threshold from the Area-Under Precision Recall Curve (AUPRC) which results in the maximum F-$\beta$ score. A discussion on the evaluation metrics used can be found in Section~\ref{sec:experiments} as well as the results pertaining to change of this threshold can be found in Figure~\ref{fig:combination}.

\subsection{Combining classification with anomaly detection}
The final consideration when constructing \ROAD~is how to effectively combine the fully supervised classifier $y_{\text{sup}} \in [0,N]$ and the fine-tuned anomaly detector $y_{\text{ssl}} \in [0,1]$. Simply, we consider normal predictions from the detector more likely to be correct, and if there is a disagreement between the two models then we flag the sample as an unknown class of anomalies that the classifier may have not seen. The overall method is shown in Figure~\ref{fig:road_diagram} and is summarised by

\begin{equation}
    y = 
    \begin{cases}
          0 &, \text{if }y_{\text{ssl}}= 0\\
          y_{\text{sup}}&,\text{if }y_{\text{ssl}}= 1\text{ and } y_{\text{sup}} \neq 0\\
          N+1 &,\text{if }y_{\text{ssl}}= 1\text{ and } y_{\text{sup}} = 0\\
    \end{cases}
    \label{eq:combine}
\end{equation}

We validate this approach in Section~\ref{sec:experiments} by showing that it is optimal assuming that normality is better defined by the SSL output.

    \begin{table*}
    \centering
    \resizebox{\linewidth}{!}{
    \begin{tabular}{|l|c|c|c|c|c|c|}
         \hline
         \textbf{Class} &  \textbf{Supervised} & \textbf{Random init} & \textbf{VAE} & \textbf{ImageNet}& \textbf{ROAD-KNN} & \textbf{ROAD} \\ 
         \hline
         Normal & \textbf{0.942 $\pm$ 0.002} &  0.940 $\pm$ 0.012 &  0.940 $\pm$ 0.004 & 0.921 $\pm$ 0.050 & 0.940 $\pm$ 0.000 & \textbf{0.942 $\pm$ 0.006} \\
         \hline
         First order data loss & 0.978 $\pm$ 0.049 &  0.975 $\pm$ 0.062 &  0.985 $\pm$ 0.031 & 0.978 $\pm$ 0.040 &  \textbf{0.988 $\pm$ 0.018} & 0.986 $\pm$ 0.039\\
         \hline
         Second order data loss & 0.767 $\pm$ 0.090 & 0.714 $\pm$ 0.122 &  0.763 $\pm$ 0.099 & 0.779 $\pm$ 0.13 &  0.726 $\pm$ 0.136 &  \textbf{0.789 $\pm$ 0.073} \\
         \hline
         High noise element & 0.777 $\pm$ 0.140 & 0.583 $\pm$ 0.205 & 0.501 $\pm$ 0.243 & 0.662 $\pm$ 0.258 &  0.719 $\pm$ 0.187 & \textbf{0.810 $\pm$ 0.162} \\
         \hline
         Oscillating tile & 0.837 $\pm$ 0.143 & 0.705 $\pm$ 0.198 &   0.752 $\pm$ 0.168 & 0.731 $\pm$ 0.198 & \textbf{0.838 $\pm$ 0.140} & 0.820 $\pm$ 0.165 \\
         \hline
         Source in the side-lobes & 0.775 $\pm$ 0.031 & 0.781 $\pm$ 0.058 & 0.741 $\pm$ 0.099 & 0.770 $\pm$ 0.076 & 0.788 $\pm$ 0.043 & \textbf{0.791 $\pm$ 0.037} \\
         \hline
         Galactic plane & 0.797 $\pm$ 0.078 & 0.781 $\pm$ 0.041 &  0.773 $\pm$ 0.053 & \textbf{0.839 $\pm$ 0.053} &  0.787 $\pm$ 0.075 & 0.819 $\pm$ 0.084\\
         \hline
         Solar storm & 0.987 $\pm$ 0.016 &  0.980 $\pm$ 0.021 &   \textbf{0.991 $\pm$ 0.028} &  0.985 $\pm$ 0.017 &  0.986 $\pm$ 0.016 & 0.987 $\pm$ 0.016\\
         \hline
         Lightning storm & \textbf{0.948 $\pm$ 0.040} & 0.897 $\pm$ 0.057 &  0.918 $\pm$ 0.060 & 0.935 $\pm$ 0.053 &  0.946 $\pm$ 0.046 &  \textbf{0.948 $\pm$ 0.042} \\
         \hline
         Ionospheric RFI reflections & 0.998 $\pm$ 0.009 &  \textbf{1.000 $\pm$ 0.000} &  \textbf{1.000 $\pm$ 0.000} & \textbf{1.000 $\pm$ 0.000} & 0.996 $\pm$ 0.012 & 0.998 $\pm$ 0.009 \\
         \hline
         
    \end{tabular}}
    \caption{F2-score classification performance on the \ROAD~dataset, where bold is the best performance per class and the ResNet34 is used for all relevant backbones.}
    \label{tab:results}
\end{table*}
\section{Experiments}
\label{sec:experiments}

We evaluate the performance of \ROAD~using the dataset described in Section~\ref{sec:dataset}. The evaluation considers both the computation and model performance using both the binary anomaly detection as well as the multi-class classification results. In all cases we use the F-$\beta$ score to evaluate the model performance. The F-$\beta$ score is the harmonic mean between precision and recall, in the context of this work precision is the anomaly detection performance that is sensitive to the number of false positives and recall is the detection performance relative to the number of false negatives. Moreover, in the context of telescope operations it is necessary to minimise the number of false negatives. In other words, it is more acceptable to classify some normal samples as anomalous than classifying anomalous samples as normal. Following this logic and work by \cite{Kerrigan2019}, we consider $\beta=2$ to be the most appropriate as it weights recall more heavily than precision. For all evaluations we use the threshold from the Area Under Precision Recall Curve (AUPRC) which maximises the F-2 score.

\subsection{Model parameters and training}
To validate our approach we experiment with several modern machine learning architectures with various model sizes. In all cases we use the same backbone architecture for both the supervised-classifier and the SSL models, furthermore, we utilise the same 2-layer MLP for position classification. Additionally, the decoder used for the SSL-reconstruction loss is a 5-layer architecture with strided de-convolution and batch-normalisation. 

For every experiment each model is trained 3 times while randomising input seeds on each run. As already mentioned in Section~\ref{sec:subsample} the low occurrence rates of some anomalous features, means we need to sub-sample the anomalous classes in the test data to ensure comparable occurrences relative to normal LOFAR telescope operations. This means we run 10 separate evaluation loops for the sub-sampled test data. In effect, the results shown in this section reflect the mean and standard deviations from 30 runs of each model. The SSL and the supervised models are trained for 100 epochs (the number of times the model is exposed to the full training set) while fine-tuning using the 2-layer MLP is done for only 20 epochs to prevent over-fitting. We use a batch size, patch size and latent dimensionality of 64 across all experiments utilising the Adam optimiser with a learning rate of $1\times 10^{-3}$ to maintain consistency. In all cases we use the official \texttt{pytorch} based implementations of the various backbones, with the exception of ViT, where we utilise an open source implementation. The code, experiments and model weights are available online\footnote{\href{https://github.com/mesarcik/ROAD}{https://github.com/mesarcik/ROAD}}.

Furthermore, to ensure no vanishing or exploding gradients while training we clip each autocorrelation to the $1^\text{st}$ and $99^\text{th}$ percentiles and take its natural log. Additionally, we normalise each magnitude-based autocorrelation between 0 and 1.

\subsection{Anomaly detection and classification}
\begin{figure}
    \centering
    \includegraphics[width=\linewidth]{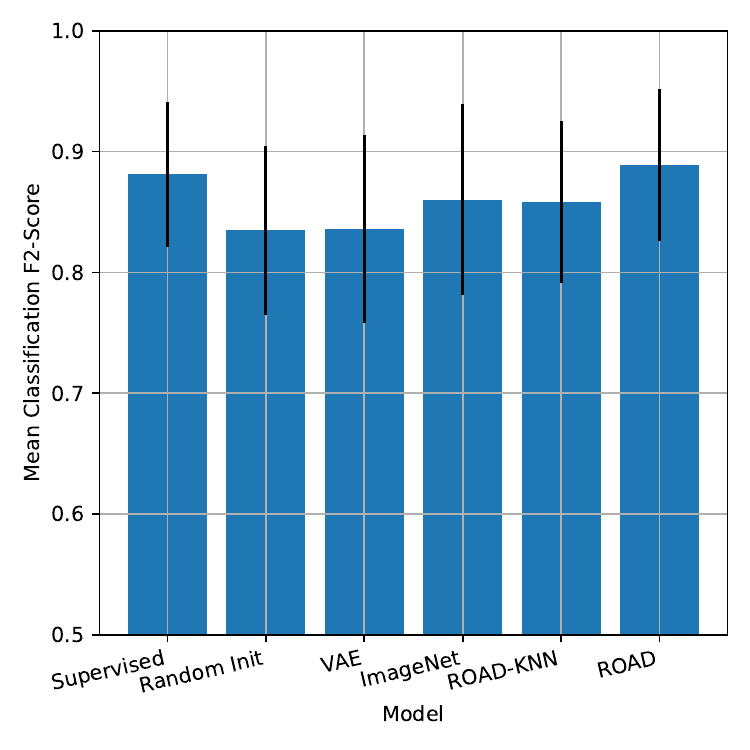}
    \caption{Per-class mean F-2 score based performance of each model shown in Table~\ref{tab:results}.}
    \label{fig:means}
\end{figure}

To maximise the model performance relative to the problem specification shown in Equation~\ref{eq:problem} we find the best mean performance of several different backbones. These being different sized ResNet~\cite{He2016}, ConvNeXt~\cite{Liu2022} and ViT~\cite{Dosovitskiy2021}. Notably our method is agnostic to backbone and could easily be extended to include architectures/model sizes. In Table~\ref{tab:results} we present the per-class results after applying the combination of the supervised-classifier and the fine tuned anomaly detector specified by Equation~\ref{eq:combine}. Furthermore, we plot the mean performance of each model in Figure~\ref{fig:means} for the sake of easy comparison. We note that all evaluated anomaly detection models  utilised fine-tuning in order to ensure they had been exposed to the same amount of data. Additionally, \ROAD-KNN utilises a KNN lookup to determine the distances in the latent space rather than using the MLP prediction. 

We find that the ResNet34 exhibits overall best average performance on the classification task, giving an average increase in F-2 score of 1\% relative to the purely supervised model. We note that the performance of \ROAD~is directly dependant on the supervised performance. We show that the SSL-pretraining is highly influential to the overall model performance as it gives a $<5\%$ increase over the randomly initialised (random init) model without pretraining. Furthermore we find that our SSL-based approach outperforms variational autoencoder-based model with fine-tuning (VAE) by $<5\%$ as well being as $<3\%$ better than KNN-based anomaly detectors (ROAD-KNN). Finally we show that using pretrained weights from ImageNet classification with fine-tuning (ImageNet) results in a $2\%$ decrease in  performance relative to our SSL pretraining paradigm.  

Across all experiments it is clear that the high noise element and oscillating tile classes have the highest standard deviation. We attribute this to the small number of examples present in both the testing and training set after adjusting for occurrence rates. In addition to this, the features represented in these classes can vary significantly from sample to sample and band to band. 

\begin{figure}
    \centering
    \includegraphics[width=\linewidth]{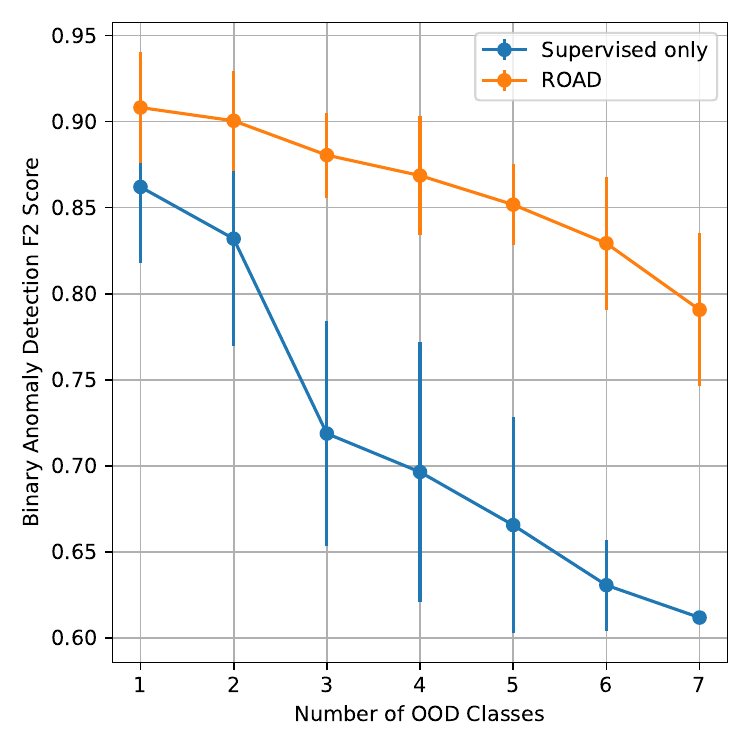}
    \caption{One-class anomaly detection performance for a purely supervised model and the fine-tuned SSL anomaly detector when removing a number of classes from the training set. The ResNet34 backbone is used for both training paradigms.}
    \label{fig:ood}
\end{figure}

\begin{figure}
    \centering
    \includegraphics[width=\linewidth]{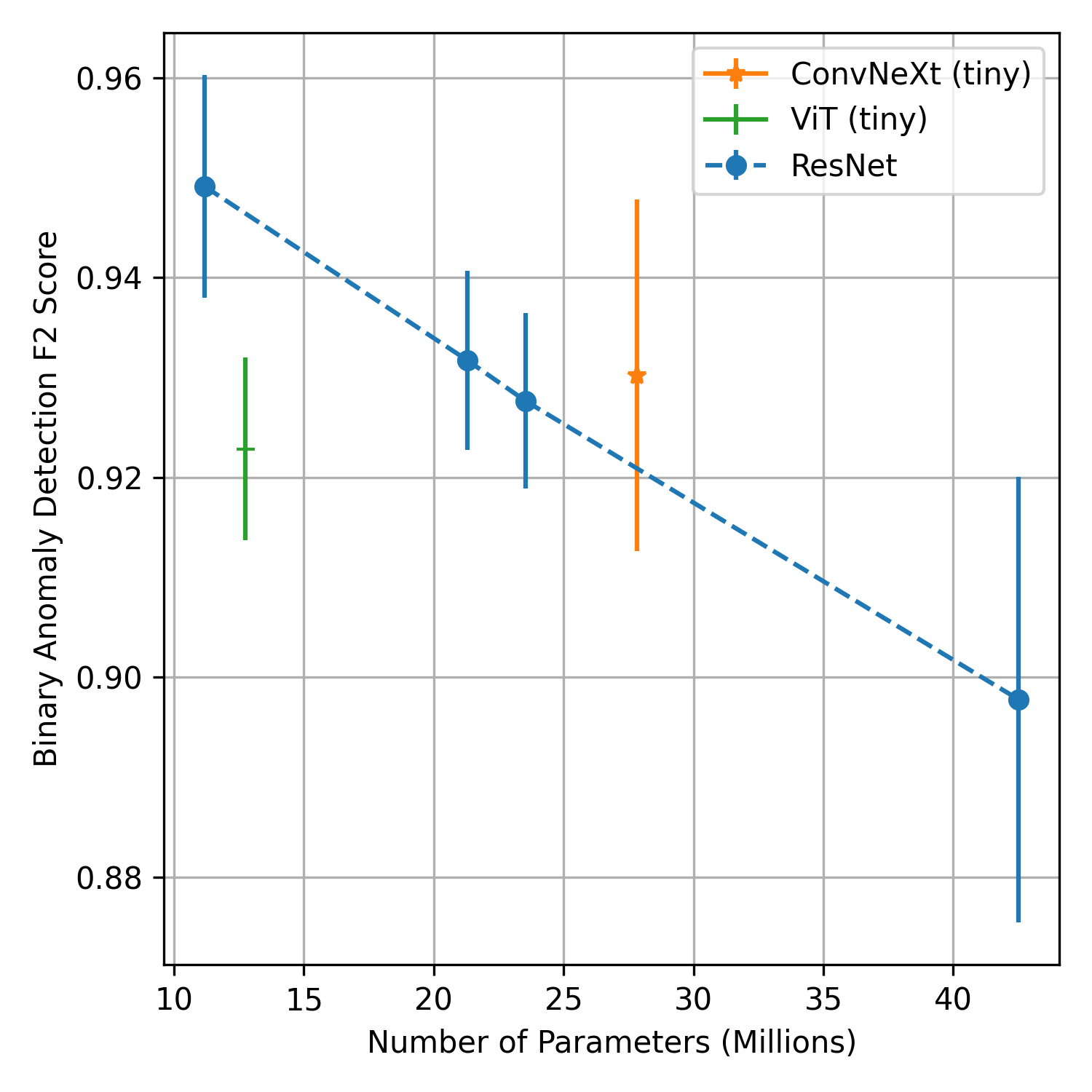}
    \caption{One-class anomaly detection performance after fine-tuning of various backbone networks when varying the number of available parameters.}
    \label{fig:backbones}
\end{figure}

\begin{figure*}[h!]
    \centering
      \includegraphics[trim={0.0cm 0cm 0cm 0cm},clip,width=\textwidth]{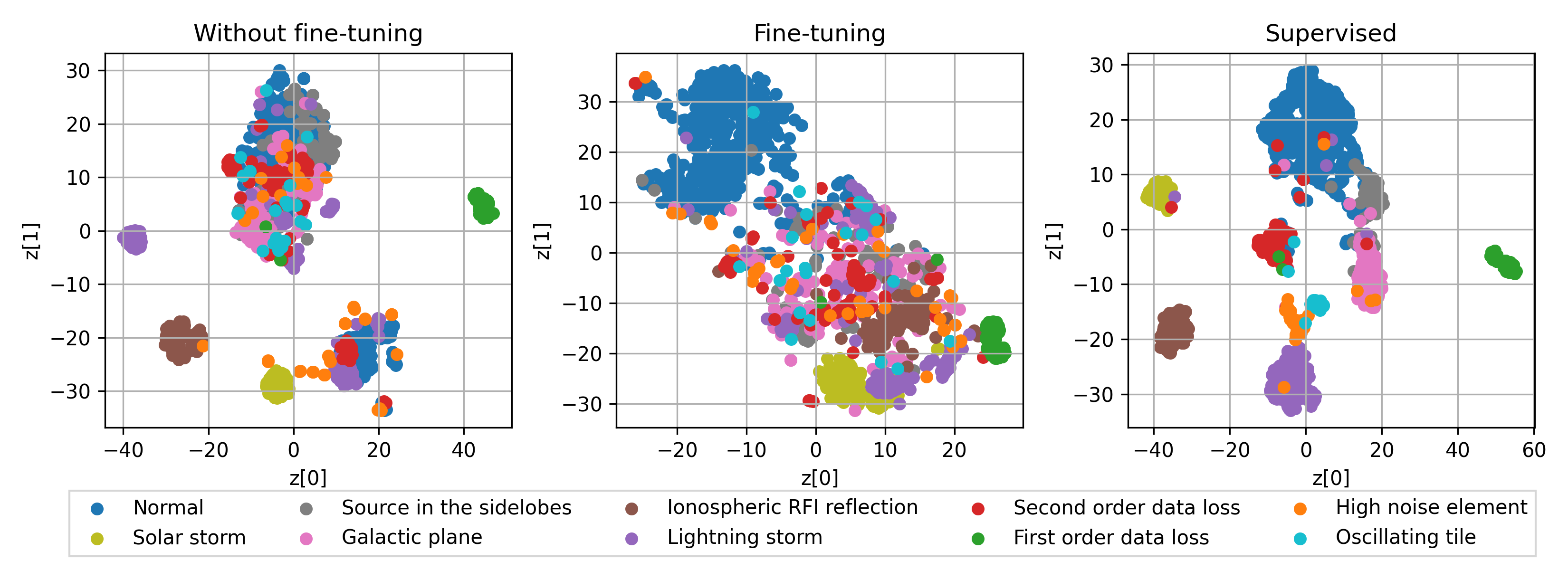}
    \caption{t-SNE projections of test data from the \ROAD~dataset using the representation from the final layer of the SSL-pretrained ResNet-34 with and without fine-tuning as well as the supervised classifier.}
    \label{fig:embeddings}
\end{figure*}

To simulate a real-world setting where many unknown anomalies can be present in a given observation, we remove several classes from the training set and test models' performance on the original test set. We refer these classes removed as \textit{Out-Of-Distribution} (OOD). The objective of this experiment is to see how well the model will react to OOD anomalies and whether it can correctly classify them as anomalous. To effectively simulate this scenario, we randomly remove between 1 and 7 classes and do this 10 times while training a model for each removal step. Figure~\ref{fig:ood} shows the average model performance from the 10 runs for both the supervised classifier as well as the fine-tuned SSL anomaly detector when removing a number of classes from the training set. Here it is clear that the supervised model suffers much more strongly from the OOD effects than the SSL-pretrained one, exhibiting a performance drop between 5\% and 18\%. Thereby illustrating the benefit of using \ROAD~where both a classifier and detector are in the loop.

We illustrate the t-distributed stochastic neighbour embedding (t-SNE) projections of the latent dimensions from each models in Figure~\ref{fig:embeddings} to gain an intuition about the model performance. The same random seed and perplexity parameters are used for all plots shown, here the perplexity estimates the number of neighbours each point should have (for more information see \cite{Wattenberg2016}). In the leftmost plot the non-fine tuned SSL-model is shown, we can see that both normal and anomalous classes are grouped closely together, with the exception of clusters pertaining to \textit{first order data-loss}, \textit{ionospheric RFI reflections} and \textit{solar storms}. Furthermore, we find the normal data is distributed across two clusters, these being LBA and HBA features. It is interesting that even with no explicit training signals the SSL model without fine-tuning is still capable of distinguishing a variety of classes and phenomena. The middle plot shows the effects of fine-tuning on the SSL representations. The fine-tuned SSL-model is significantly better at distinguishing normal from anomalous samples, with the LBA/HBA separation in the normal samples completely disappearing. Furthermore, the clusters corresponding to features that were once well separated such as \textit{solar storm} are now better grouped with the anomalous samples. Finally in the right-most plot we can see the learn supervised representations of the test data. Here it is clear that the supervised model is the most capable of separating both anomalous and normal classes alike. It must be noted however that the classes relating to \textit{galactic plane}, \textit{source in the sidelobes} and \textit{normal} are overlapping. Therefore by combining the boundary related to the SSL-fine-tuned embedding with the specificity of the supervised model we are able to better detect anomalies.

\begin{figure}
    \centering
    \includegraphics[width=\linewidth]{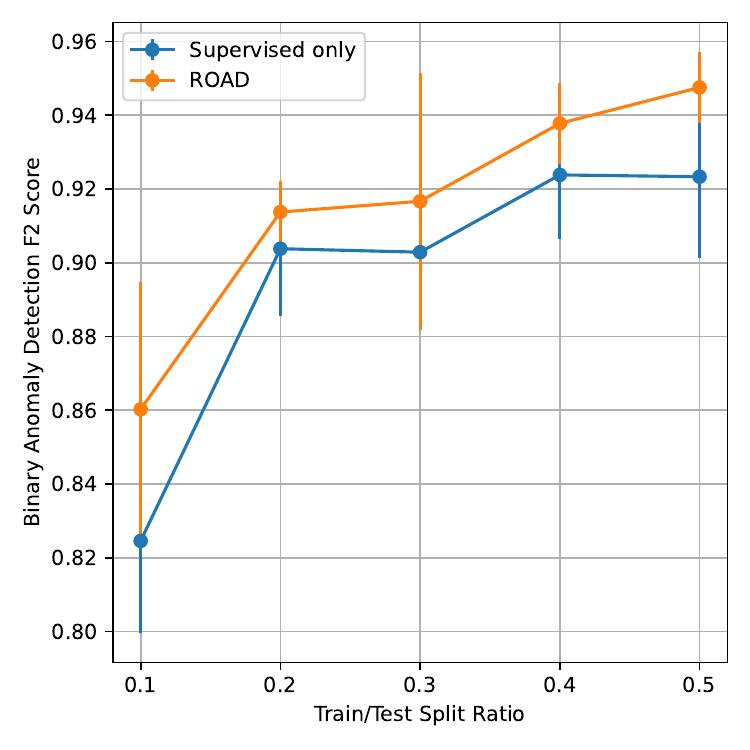}
    \caption{Binary anomaly detection performance when changing the amount of supervision used for training a ResNet-34 backbone for each training paradigm.}
    \label{fig:percentage}
\end{figure}

An interesting consequence of the class imbalance and the few number of samples certain events such as \textit{oscillating tile}, is that \ROAD~benefits from fewer backbone parameters and does not scale with model size, as it over-fits to the training data. This is illustrated in Figure~\ref{fig:backbones}, where it is also shown that ResNets offer the best performance. This being said, we expect that with more samples from the infrequent classes the model performance should scale proportionally with its number of parameters. This is further validated by Figure~\ref{fig:percentage}, where we plot the model performance relative to the amount of training data. Here is is clear that the model performance scales linearly with training data-size. Furthermore, the fine-tuned model outperforms its purely supervised counterpart for all training set sizes.

\begin{figure*}
    \centering
    \includegraphics[width=\linewidth]{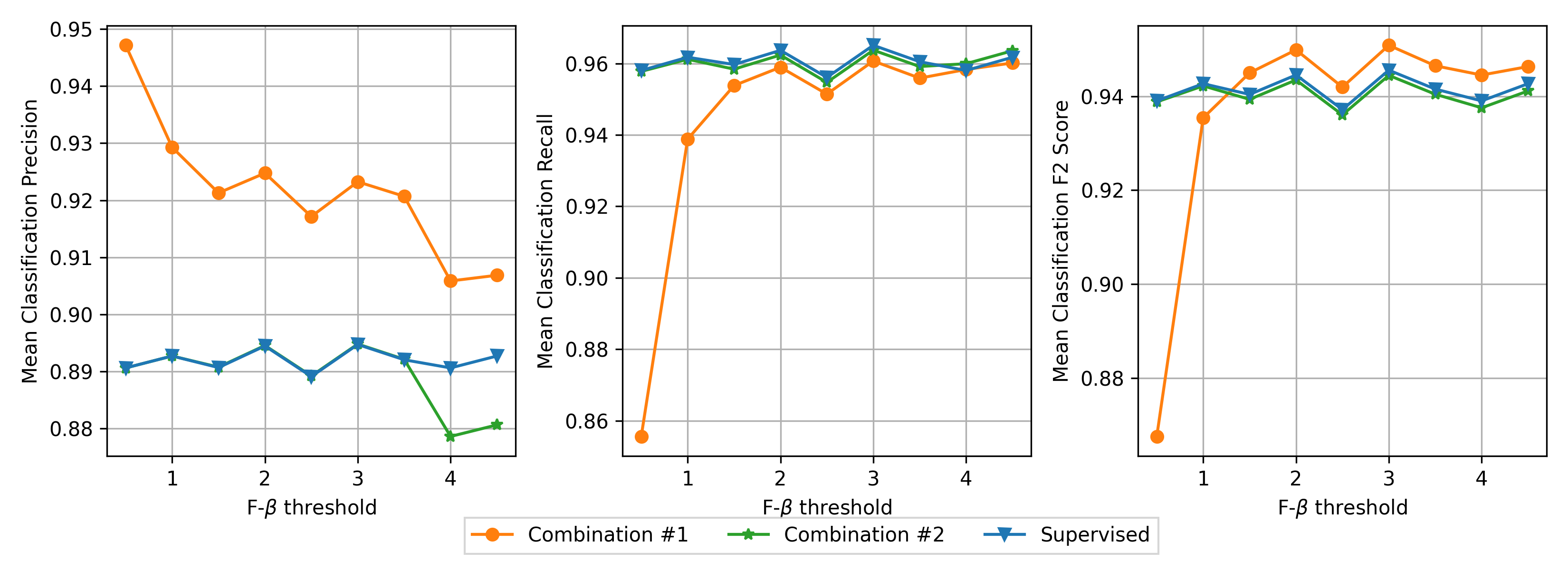}
    \caption{Mean classification performance of the ResNet-34 backbone after fine-tuning when changing the threshold used for anomaly detection as well as the combination function. Combinations \#1 and \#2 correspond to Equations~\ref{eq:combine} and \ref{eq:combination_2} respectively. }
    \label{fig:combination}
\end{figure*}

\subsection{Model ablations}
 To validate the correctness of the SSL-model training objective we perform several ablations. In Table~\ref{tab:ablations} we show the effect of only using only the reconstruction term, $\mathcal{L}_{\text{recon}}$, or only the context prediction term, $\mathcal{L}_{\text{con}}$, or using the combined loss $\mathcal{L}_{\text{recon}} + \mathcal{L}_{\text{con}}$. We show that the combination of the two terms improves both the anomaly detection and the average classification performances by $~2\%$, which at the scale of the LOFAR science data processing pipeline results in a significant improvement.

 \begin{table}[h!]
    \centering
    \resizebox{\linewidth}{!}{
    \begin{tabular}{|l|c|c|c|}
         \hline
         \textbf{Performance} & \textbf{$\mathcal{L}_{\text{recon}}$} & \textbf{$\mathcal{L}_{\text{con}}$} & 
    \textbf{$\mathcal{L}_{\text{recon}}+\mathcal{L}_{\text{con}}$} \\
         \hline
         Anomaly detection &  0.88 $\pm$ 0.03  & 0.92 $\pm$ 0.01 & \textbf{0.93 $\pm$ 0.01} \\
         \hline
         Classification & 0.85 $\pm$ 0.07 & 0.87 $\pm$ 0.05 &\textbf{0.89 $\pm$ 0.06}\\
         \hline
         
    \end{tabular}}
    \caption{Model performance (F2 score) after fine-tuning when varying the SSL loss function for a ResNet34 backbone.}
    \label{tab:ablations}
\end{table}

\begin{figure}
    \centering
    \includegraphics[width=\linewidth]{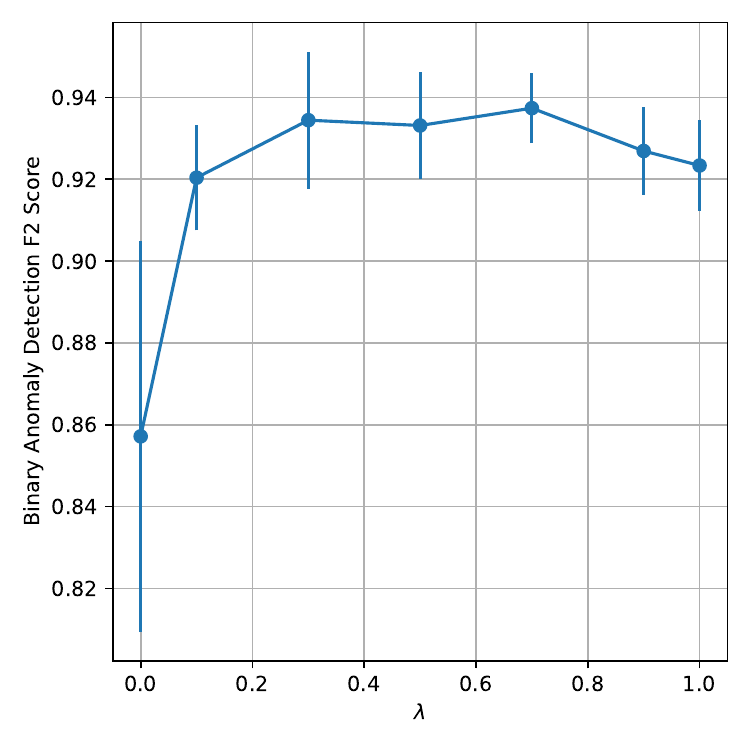}
    \caption{ Mean anomaly detection F2-score performance of the ResNet-34 when varying the relative contribution of $\mathcal{L}_{\text{recon}}$ and $\mathcal{L}_{\text{con}}$. When $\lambda$ is minimum the loss favours $\mathcal{L}_{\text{recon}}$ while a for greater losses $\mathcal{L}_{\text{con}}$ dominates.}
    \label{fig:lambdas}
\end{figure}

Furthermore, in order to determine the relative contribution of each of the losses to the overall performance of \ROAD~we modify the $\lambda$ hyper-parameter and measure the overal model performance. Figure~\ref{fig:lambdas} shows how with $0.3\leq \lambda \leq 0.7$ the SSL anomaly detection obtains optimal performance.

In addition to the loss function-based ablations we also consider the effect of changing the combination function used between the supervised and SSL model shown in Equation~\ref{eq:combine}. These results are shown in Figure~\ref{fig:combination}, when we vary both the anomaly detection threshold set by the maximum F-$\beta$ score as well as the combination function. In the plot, \textit{combination function \#1} uses the definition expressed in Equation~\ref{eq:combine}, where the anomaly detector defines both normality and the unknown anomaly events. We define \textit{combination function \#2} as 
\begin{equation}
    y =\begin{cases}
          N+1 &,\text{if }y_{\text{ssl}}= 1\text{ and } y_{\text{sup}} = 0\\
          y_{\text{sup}} &,\text{otherwise }
    \end{cases}
    \label{eq:combination_2}
\end{equation}
such that $y_{\text{ssl}}$ is only used to define unknown anomalous events. In the leftmost plot we can see that  combination function \#1 consistently offers the best precision, while at the cost of a marginally decreasing the recall (<0.4\%). The effect of this is that combination function \#1 results in optimal F-2 score performance when the $\beta$ is greater than 1. Futhermore, we evaluate the false positive rate using combination function \#1 and find that it results in a false positive rate of approximately 2\%.

\subsection{Computation performance analysis}
We evaluate the computational performance of \ROAD~during inference on a Nvidia A10 GPU using \texttt{CUDA}~11.7 and using driver release \texttt{515.65.01}. The KNN-based experiments utilise the GPU-based implementation of \texttt{FAISS}\footnote{\href{https://github.com/facebookresearch/faiss}{https://github.com/facebookresearch/faiss}}. We use a batch size of 1024, with a patch size and a latent dimensionality of 64. Furthermore for the case of the KNN search we assume 1000 normal training samples to populate the search space. In all cases we use \texttt{bfloat16} representations of the input data such that to ensure the tensor-cores are fully utilised. In these results, we perform 1000 forward passes and measure the resulting latency, throughput in spectrograms per second as well as peak memory allocation. 

The computation performance of the respective models can be seen in Table~\ref{tab:performance}, where it is clear that the supervised model has the lowest computational overhead. We relate the difference performance between the supervised and SSL model to the dimensionality of the models inputs and required concatenation of the patches on each forward pass. As the SSL operates on the patch level, there are substantially fewer convolution operations that need to be applied (approximately 16), resulting in decreased peak memory performance. \ROAD~consists of both the supervised and SSL models and such the overall performance is given by the addition of the respective values, such that it takes less than 1~ms to predict the normality of a given spectrogram. This is more than 1000x faster than the existing correlator implementations on the  IBM Blue Gene/P supercomputer~\cite{Romein2010}. Notably however, the KNN-based model performs significantly worse, suggesting that density based KNN anomaly detectors are less suitable for real-time applications at observatories. 

\begin{table}[h!]
    \centering
    \resizebox{\linewidth}{!}{
    \begin{tabular}{|l|c|c|c|c|}
         \hline
         \textbf{Model Type} & \textbf{Latency (ms)} & \textbf{Throughput(spec/s)}& \textbf{Peak Memory(GBs)} \\
         \hline
         Supervised & 0.3 & 2844 & 8.11 \\
         \hline
         SSL & 0.4 & 2481 & 5.95 \\
         \hline
         KNN & 12.1 & 82 & 5.95 \\
         \hline
    \end{tabular}}
    \caption{Computational performance of anomaly detectors, where spec/s referes to the number of spectrograms processed per second by the respective algorithm.}
    \label{tab:performance}
\end{table}

    \section{Conclusions and future work}
\label{sec:conclusions}
In this work we have presented the first real time anomaly detector for system-wide anomalies in spectrographic data from radio telescopes. We produced a freely available dataset that contains 7050 autocorrelation-based spectrograms from the LOFAR telescope with labels relating to both commonly occurring anomalies as well as rare events. This work provides a formulation of anomaly detection in the SHM-context of telescope operations and illustrates how purely supervised models are ill-suited to the problem. Furthermore, we propose a new Self-Supervised Learning (SSL) paradigm for learning normal representations of spectrographic data. We combine both the SSL and supervised models and demonstrate how it remedies the shortcomings of supervised methods. We demonstrated that even with limited examples of anomalous data our fine tuned SSL model can significantly outperform its supervised counterpart.  The radio observatory anomaly detector (ROAD) and dataset are the first major effort to address the system health management problem in radio telescopes and its potential benefit to all radio observatories is very promising.  

We expect through providing open source access to both our models and dataset, continued effort by the larger community will increase the amount of training data from scarce events. Thereby enabling other training paradigms such as contrastive learning with larger models that are currently unsuited to the highly imbalanced problem. Furthermore, we identify several directions for future work in the area of radio observatory anomaly detection. Namely using the cross correlations to enhance training by using radio interferometer specific losses. Another interesting direction would be to use Bayesian deep learning to give uncertainty estimates from the classifier such that samples with low confidence would rely on the detector output. Finally, we would like to propagate the labels from the down-sampled data to the full resolution data from LOFAR Long Term Archive such that the performance could be better evaluated in the context of the full LOFAR data processing pipeline.
    \begin{acknowledgements}
This work is part of the "Perspectief" research programme "Efficient Deep Learning" (EDL, \href{https://efficientdeeplearning.nl}{https://efficientdeeplearning.nl}), which is financed by the Dutch Research Council (NWO) domain Applied and Engineering Sciences (TTW). The research makes use of  radio astronomy data from the LOFAR telescope, which is operated by ASTRON (Netherlands Institute for Radio Astronomy), an institute belonging to the Netherlands Foundation for Scientific Research (NWO-I). 
\end{acknowledgements}
    \bibliography{references}
    \begin{appendix} 
\end{appendix}
\end{document}